\RequirePackage{fix-cm}

\documentclass[preprint,smallcondensed]{svjour3}

\smartqed

\usepackage[english]{babel}

\usepackage{graphicx}
\usepackage{latexsym}
\usepackage{natbib}
\usepackage[tight,TABTOPCAP]{subfigure}
\usepackage[colorlinks=true,citecolor=blue,urlcolor=blue]{hyperref}
\usepackage{url}
\usepackage{grffile}% usefull: avoid error with point in name figure
\usepackage{rotating}
\usepackage{multirow}
\usepackage{microtype}
\usepackage{color}
\usepackage{amstext,amssymb,amsmath}
\usepackage{tikz}
\usepackage{siunitx}
\usepackage{placeins}
\usepackage{xcolor} 
\newsavebox{\largestimage}
\newcommand{\blackline}{\raisebox{2pt}{\tikz{\draw[-,black!,solid,line width = 0.9pt](0,0) -- (5mm,0);}}\;\;}
\newcommand{\redline}{\raisebox{2pt}{\tikz{\draw[-,red!,solid,line width = 0.9pt](0,0) -- (5mm,0);}}\;\;}
\newcommand{\grayline}{\raisebox{2pt}{\tikz{\draw[-,gray!,solid,line width = 0.9pt](0,0) -- (5mm,0);}}\;\;}
\newcommand{\blueline}{\raisebox{2pt}{\tikz{\draw[-,blue!,solid,line width = 0.9pt](0,0) -- (5mm,0);}}\;\;}
\newcommand{\greenline}{\raisebox{2pt}{\tikz{\draw[-,green!,solid,line width = 0.9pt](0,0) -- (5mm,0);}}\;\;}

\newcommand{\dashedgrayline}{\raisebox{2pt}{\tikz{\draw[-,gray!,dashed,line width = 0.9pt](0,0) -- (5mm,0);}}\;\;}

\newcommand{\dottedblackline}{\raisebox{2pt}{\tikz{\draw[-,black!,dotted,line width = 0.9pt](0,0) -- (5mm,0);}}\;\;}

\definecolor{orcidlogocol}{HTML}{A6CE39}
\definecolor{dgreen}{rgb}{0.2,0.5,0.2}

 \usepackage[authormarkup=none,final]{changes}
  \definechangesauthor[name={Reviewer 1},color=red]{1}
  \definechangesauthor[name={Reviewer 2},color=blue]{2}
  \definechangesauthor[name={Reviewer 3},color=dgreen]{3}
  \definechangesauthor[name={Author 4},color=orange]{4}
 \setauthormarkup{}

\definecolor{burntorange}{rgb}{0.8, 0.33, 0.0}

\definecolor{cadmiumgreen}{rgb}{0.0, 0.42, 0.24}

\DeclareGraphicsExtensions{.pdf,.png}

\graphicspath{{images/}}

\begin{document}

%%%%%%%%%%%%%%%%%%%%%%%%%%%%%%%%%%%%%%%%%%%%%%%%%%%%%%%%%%%%%%%%%%%%%%%%%%%%%%%%%%

\title{Numerical investigation of the boundary layer stability on a section of a rotating wind turbine blade}

\titlerunning{Investigation of the boundary layer stability on a rotating wind turbine blade}

\journalname{Flow, Turbulence and Combustion}      

\author{T.\ C.\ L.\ Fava \and D.\ S.\ Henningson \and A.\ Hanifi}
\authorrunning{T C L Fava et al.}    

\institute{T C L Fava (Corresponding Author) \and D.\ S.\ Henningson \and A.\ Hanifi \at 
                 KTH Royal Institute of Technology, Dept.\ of Engineering Mechanics, FLOW and SeRC, SE-100 44 Stockholm, Sweden,
                 \email{fava@kth.se, henning@mech.kth.se, hanifi@kth.se}}

\date{31 July 2023}

\maketitle

%%%%%%%%%%%%%%%%%%%%%%%%%%%%%%%%%%%%%%%%%%%%%%%%%%%%%%%%%%%%%%%%%%%%%%%%%%%%%%%%
%%%%%%%%%%%%%%%%%%%%%%%%%%%%% --- ABSTRACT --- %%%%%%%%%%%%%%%%%%%%%%%%%%%%%%%%%
%%%%%%%%%%%%%%%%%%%%%%%%%%%%%%%%%%%%%%%%%%%%%%%%%%%%%%%%%%%%%%%%%%%%%%%%%%%%%%%%

\begin{abstract}
{
Direct numerical simulations (DNS) and linear stability analysis of the flow on a rotating wind turbine blade section are performed to study the effects of rotation on laminar-turbulent transition on the suction side of the airfoil. A chord Reynolds number of $1 \times 10^5$ and angles of attack ($AoA$) of $12.8^\circ$, $4.2^\circ$, and $1.2^\circ$ were considered. The first set of simulations considered Coriolis and centrifugal forces that appear in the rotating frame of reference. The second one disregarded these effects, but the same angle of attack and Reynolds number were kept. Rotation effects on the flow depend on the streamwise pressure gradient and direct/reverse flow state. Upon separation in the $AoA=12.8^\circ$ case, rotation retarded the flow and rendered the mixed Tollmien-Schlichting/Kelvin-Helmholtz (TS/KH) instability in the laminar separation bubble (LSB) more unstable. Moreover, an oblique secondary instability mechanism was more intense in the rotating case, leading to a rapid breakdown of the rolls to small-scale turbulence. On the other hand, a sub-harmonic mechanism was dominant in the non-rotating case, and the rolls required a longer streamwise extent to lose spanwise coherence. Despite that, rotation accelerated the attached boundary layer upstream of the LSB, which was subjected to a strong adverse pressure gradient (APG), stabilizing TS waves and rendering the transition location 3\% more downstream in the rotating case. The reattachment location was shifted 4\% downstream in the rotating case, which was both an effect of transition delay and flow deceleration by rotation. In the  $AoA=4.2^\circ$ and $AoA=1.2^\circ$ cases, rotation decelerated the flow upstream of separation, where a favorable pressure gradient (FPG) is present. This increased the growth rates of TS waves in this region. However, rotation slightly accelerated the flow inside the LSB, rendering the TS/KH mode less unstable. Furthermore, the high spanwise velocity and inflection crossflow profiles led to the appearance of stationary and traveling crossflow modes in these cases, which enhanced the perturbation level in regions of tip flow. Nonetheless, these modes presented low energy compared to the TS/KH mechanism in the separated shear layer, which remained the dominant instability. An effect of the crossflow was to generate coherence in the spanwise component of the TS/KH mechanism, where a wavepacket was formed close to the location of the inflection point in the spanwise velocity profiles. Moreover, the secondary instability mechanism was mostly oblique in the non-rotating case for $AoA=4.2^\circ$, whereas it was mainly sub-harmonic in the rotating one. In the $AoA=1.2^\circ$ case, the sub-harmonic mechanism seemed to dominate both conditions. Modes with negative spanwise wavenumber were preferentially excited in the rotating case. In both $AoA=4.2^\circ$ and $AoA=1.2^\circ$ cases, these effects did not affect the transition location. A finite region of absolute instability was found in these cases, where rotation generally rendered this mechanism less unstable.
}

\bigskip 
\bigskip 
%% Keywords must be edited
\end{abstract}

%%%%%%%%%%%%%%%%%%%%%%%%%%%%%%%%%%%%%%%%%%%%%%%%%%%%%%%%%%%%%%%%%%%%%%%%%%%%%%%%%
%%%%%%%%%%%%%%%%%%%% --- 1 INTRODUCTION --- %%%%%%%%%%%%%%%%%%%%%%%%%%%%%%%%%%%%%
%%%%%%%%%%%%%%%%%%%%%%%%%%%%%%%%%%%%%%%%%%%%%%%%%%%%%%%%%%%%%%%%%%%%%%%%%%%%%%%%%

\section{Introduction}
\label{sec:introduction}

In a pioneering study, \citet{himmelskamp1947} noted that rotating aircraft propellers presented modified aerodynamic characteristics compared to static blades, particularly a higher lift, a phenomenon later called rotational augmentation. Nevertheless, early theoretical investigations with a blade modeled as a flat plate indicated reduced rotation influence on the attached flow region \citep{fogarty1951,mccroskey1971}. The study of a helical blade by  \citet{banks1963} demonstrated that rotation delays or even suppresses separation, especially in the inboard region, if the adverse pressure gradient (APG) is low enough. \citet{horlock1965} showed that the crossflow strongly depends on rotation, and \citet{mccroskey1968} found that this velocity component is crucial for the separation delay. This phenomenon draws, in particular, the attention of the wind-energy community due to the important flow separation over the blades, often used to prevent over-speed in the so-called stall-regulated wind turbines \citep{corten2001}. In this regard, \citet{savino1985} and \citet{bosschers1995} confirmed the downstream moving of the separation point experimentally for increasing rotation rates. \citet{du2000} solved the integral momentum boundary layer equations considering rotation and
also found separation and stall delays and lift enhancement for decreasing Reynolds ($Re$) and increasing rotation ($Ro_r$) numbers and solidity ($c/r$). Here, $Ro_r=\Omega r/V_\infty$, where $\Omega$ is the rotation speed, $r$ is the radius, $V_\infty$ is the wind speed, and $c$ the chord length. Other works also identified the increase in solidity to promote stronger rotation effects and a direct proportionality between this parameter and the lift coefficient \citep{shen1999,chaviaropoulos2000}. Nonetheless, the latter two works, which involved quasi-two-dimensional computations of a blade section, indicated no change in the separation location but rather a pressure reduction in this region. Moreover, \citet{dumitrescu2007} indicated that the separation location was little affected by the solidity, but the reattachment point shifted upstream, and the spanwise velocity increased with $c/r$.

A consequence of these findings is that the inboard region of the blade is more affected by rotation. This zone, particularly for radial locations below 30\% of the rotor radius, presents a marked stall delay and lift increase, clearly observed in experiments \citep{ronsten1992,bjorck1995}. In regions where the balance between the Coriolis and centrifugal forces does not occur, such as in the separated flow, the latter imparts a radial velocity component to stagnant or reverse flow, known as centrifugal pumping \citep{herraez2014}. Indeed, flow visualizations of full-scale wind-turbine rotors indicated chordwise-oriented flow upstream of separation and after turbulent reattachment but strongly radial streamlines upon separation \citep{mccroskey1973,savino1985,ronsten1992,corten2001,herraez2014}, possibly reaching 80\% of the azimuthal velocity \citep{chaviaropoulos2000}. This was also noticed in experiments of rotating propeller blades \citep{schulein2012}. The Coriolis force acting on this radial flow towards the tip tends to accelerate the flow downstream, effectively acting as a favorable pressure gradient (FPG) \citep{bosschers1995,breton2008,dumitrescu2004,herraez2014}.
\citet{chaviaropoulos2000} and \citet{dumitrescu2004} concluded that 
the radial component sucks low-momentum fluid from the separation region, lowering the pressure on the suction side. This vision is also shared by \citet{schreck2007}, which concluded that the lift enhancement correlated with the magnitude of the radial velocity component.

Due to the complexity of experiments and the high computational cost of numerical investigations, investigations on laminar-turbulent transition in blades have disregarded rotation effects \citep{lang2015}. As indicated by \citet{hernandez2007}, most numerical studies assume the boundary layer to be entirely turbulent or employ inadequate or oversimplified criteria for the prediction of laminar-turbulent transition. \citet{hernandez2012} analyzed the effects of rotation on transition in a blade by generating a base flow for stability analysis using integral boundary layer equations. Rotation stabilized the flow to TS waves upstream of separation and on the pressure side. \citet{pascal2013} computed the flow over a rotating fan blade with Reynolds averaged Navier-Stokes (RANS) equations and performed local linear stability analysis of the flow profiles. Rotation almost did not change the growth rates of TS waves but destabilized stationary crossflow modes, which were too weak to trigger transition. \citet{fava2021} performed a similar analysis but considered more general disturbances and concluded that transition was shifted upstream in regions with strong rotation effects and three-dimensionality (e.g., the inboard region). Highly oblique modes were found to play a role in this phenomenon. \citet{jing2020a} performed a direct numerical simulation (DNS) of a wind-turbine section ranging from 38\% to 88\% of the rotor radius at a chord Reynolds number $Re_c=3 \times 10^5$. Inflectional spanwise velocity profiles were obtained, but they presented a maximum of 5\% of the streamwise velocity, and crossflow instability was not present. Transition occurred via Tollmien-Schlichting (TS) waves, with waves slightly misaligned with the spanwise direction. However, these simulations were performed for near-optimal conditions, reducing separation and rotation effects. \citet{gross2012} studied two sections of a rotating wind turbine blade with DNS at 20\% and 80\% of the rotor radius, with an angle of attach ($AoA=5^\circ$) and showed that rotation triggered stationary and traveling crossflow modes due to the generation of a crossflow velocity component, which contributed to an earlier transition, separation delay, and lift increase. \citet{guntur2012} indicated that the previous study lacked the analysis of higher $AoA$, more relevant for wind-turbine applications. Furthermore, \citet{gross2012} did not account for rotation terms in the stability calculations. In their implicit LES of a helicopter blade in yaw, \citet{wen2019} found that the structures shed from the LSB were oblique and concluded their origin might be crossflow instability. \citet{jing2020b} employed DNS to analyze a rotating marine propeller and found that the radial velocity can reach around 14\% (50\% near the hub upon separation) of the azimuthal velocity, triggering crossflow instability and transition to turbulence, similar to the rotating disk. In this regard, the flow was absolutely unstable in the radial direction and convectively unstable in the streamwise one. Experiments by \citet{yang2006} indicated the formation of discrete quasi-periodic streamwise vortical structures (spanwise cells) in the separated flow on a rotating hovering blade, which may indicate crossflow vortices. The PIV measurements of \citet{diottavio2008} and \citet{raghav2014} confirmed these structures. They found that despite being present in fixed wings under stall \citep{winkelmann1980,bippes1982,wehis1983,schewe2001}, the radial flow generated by rotation made them more pronounced. Since these cells could be linked with a steady three-dimensional global mode in the non-rotating case \citep{theofilis2000,rodriguez2010}, rotation may affect the global stability of such flows.

Further results on the effects of rotation on the boundary layer stability and transition are mainly restricted to canonical configurations, such as rotating channels \citep{alfredsson1989,wall2006}, flat plates \citep{potter1971,deschamps2017}, and disks \citep{malik1981,hall1986,malik1986,balachandar1992,lingwood1995,lingwood1996}. In the case of rotating disks with axial inflow, which may be a simplified model for a rotating blade, \citet{hussain2011} and \citet{deschamps2017} found that the axial inflow was stabilizing vis-à-vis crossflow instabilities since the crossflow velocity profiles lose their inflectional character. 

From the above, little is known about the
impact of rotation on transition mechanisms on rotating blades \citep{pascal2013,lang2015,deschamps2017}. This is partly due to the limited number of studies carrying out detailed numerical simulations of the transition process on rotating blades. Furthermore, the available results were performed for near-optimum operation conditions, which tends to prevent separation and mitigate the rotation effects \citep{jing2020a}. The lack of parametric change in the rotation conditions yields another knowledge gap \citep{guntur2012}. For this reason, the current work attempts to shed light on the above phenomena through the detailed DNS of a rotating wind turbine blade section at 68\% of the rotor radius, $Re_c = 1 \times 10^5$, and three operational conditions. The latter consists of three angles of attack, corresponding to three rotation numbers or tip-speed ratios. Accompanying simulations with the same parameters but without rotation are also performed for comparison. The laminar-turbulent transition process is analyzed and compared with linear stability results. The paper is divided as follows. \S \ref{sec:num_setup} presents the numerical setup, including the blade geometry, numerical method for obtaining the flow, study cases, and temporal and spatial convergence studies.  \S \ref{sec:results} displays the results. The mean-characterization is performed in \S \ref{sec:mean_flow}, and instantaneous structures are presented in \S \ref{sec:instat_flow}. Linear primary stability analysis results are shown in \S \ref{sec:stab}. Further investigation of the flow through Fourier and SPOD analysis is carried out in \S \ref{sec:spectral}. Finally, \S \ref{sec:conclusion} presents the conclusions of this work.

\section{Numerical setup}\label{sec:num_setup}

\subsection{Blade model}\label{sec:rot_blade}

In the following, unless otherwise stated, the variables are non-dimensionalized with the chord length ($c$), and the resultant free-stream velocity ($U_\infty=\sqrt{V_\infty^2+(\Omega r)^2}$) from the wind ($V_\infty$) and rotation ($\Omega r$), where $\Omega$ is the rotation speed, and $r$ is the radial position of the simulated domain.

\begin{figure}[!htb]
    \centering
    \begin{minipage}{1\textwidth}
    \centering
    \includegraphics[width=1\linewidth,trim={0cm 0cm 0cm 0cm},clip]{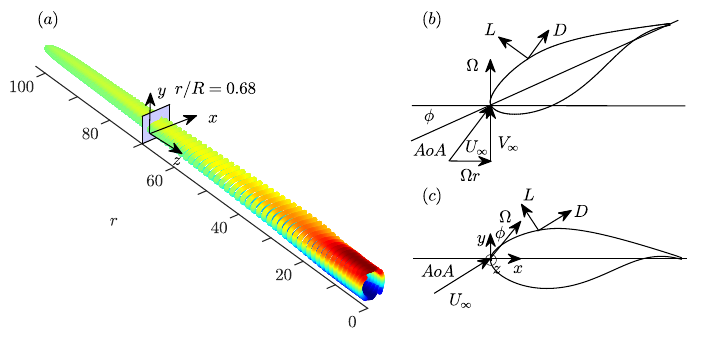}
    \caption{(a) DTU 10-MW Reference Wind Turbine blade and domain of simulation at $r/R=0.68$ ($R$ is the tip radius in meters). (b) Physical frame of reference. (c) Simulation frame of reference.}
    \label{fig:blade_r068}
    \end{minipage}
\end{figure}

The simulation domain consists of a slice of the DTU 10-MW Reference Wind Turbine \citep{bak2012} at 68\% of the rotor radius ($R$), whose profile corresponds to a blend of 96\% of FFA-W3-241, and 4\% of the FFA-W3-301 airfoils \citep{bjorck1990}. The blade and the slice simulated are shown in Fig. \ref{fig:blade_r068} (a). The domain has a span width of 10\% of the chord length, but no spanwise geometry variation is considered to allow a periodicity condition in this direction. The chord Reynolds number was $Re_c = U_\infty c/ \nu = 1 \times 10^5$, where $\nu$ is the kinematic viscosity, local solidity $c/r=0.055$, and geometric twist angle $\phi=4.8^\circ$.  Figure \ref{fig:blade_r068} (b) shows the blade cut by a cylindrical shell with the centerline in the vertical direction aligned with the rotation axis ($\Omega$ vector). $AoA$ is the angle of attack, and $L$ and $D$ are the lift and drag forces. The geometries were rotated in the clockwise direction by $\phi$, as portrayed in Fig. \ref{fig:blade_r068} (c), so that the $x$ coordinate of the simulation frame of reference lies along the airfoil chord. In this way, the $y$ and $z$ coordinates point in the vertical and out-of-plane directions, respectively.

%-------------------------------------------------------------------------------
%-------------------------------------------------------------------------------

\subsection{Numerical method}\label{sec_num_method}

The simulations were performed with the incompressible spectral-element Navier-Stokes solver Nek5000 \citep{fischer2008}. The rotating reference frame fixed to the blade was adopted for the computations. The Coriolis and centrifugal forces were included as volume sources in the momentum equations and are given by

\begin{gather}
        f_x = -2 \Omega_x u_z, \nonumber\\
        f_y = +2 \Omega_y u_z, \\
        f_z = -r \Omega^2 + 2 \Omega_y u_x - 2 \Omega_x u_y,\nonumber
\end{gather}

\noindent where $u_x$, $u_y$, and $u_z$ are the velocity components in $x$, $y$, and $z$. Initial and boundary conditions were computed with RANS simulations using the solver Fluent \citep{fluent}, where Coriolis and centrifugal forces were also included. The domain size of the RANS computations was five chord lengths in every direction. The DNS domain extends one-chord length from the airfoil in every direction except the wake, where the outlet is two-chord lengths downstream of the trailing edge. The mesh has 246,000 elements, each discretized with eight Gauss-Lobatto-Legendre (GLL) points in every direction (seventh-order polynomial approximation), yielding 126 million grid points. A time step $\Delta t = 5 \times 10^{-6}$ was used.

A uniform Dirichlet boundary condition for the velocity was imposed on the inlet of the RANS simulations, consisting of the velocity with magnitude $U_\infty$ and angle $AoA$ on the $xy$ plane. In addition, zero spanwise velocity was imposed on this boundary, which is far from the blade. A zero-pressure Neumann boundary condition was enforced on the outlet. The velocity field computed with RANS was interpolated onto the inlet surface of the DNS domain and used as a Dirichlet boundary condition. An open boundary condition where $\left[-p \mathbf{I}+Re_c^{-1}\nabla \mathbf{U}\right] \cdot \mathbf{n}=0$ was applied to the outlet ($p$ is pressure, $\mathbf{I}$ the identity matrix, $\mathbf{U}$ the velocity vector, and $\mathbf{n}$ the normal vector). Finally, periodic boundary conditions were employed in the spanwise direction of both RANS and DNS

The spectral element method is known for displaying low noise levels, which may preclude transition. In order to prevent that, the tripping line described in \citet{schlatter2012} was applied at 5\% of the chord on the suction side with an unsteady three-dimensional random forcing with an amplitude of $1\times10^{-6}$ uniformly distributed over frequency until $f=630 \; U_\infty/c$. A fringe was applied to the last 50\% of the chord closest to the outlet to promote an outflow normal to the boundary and without backflow events. The implicit filter of \citet{negi2017} was employed, albeit with a small amplitude corresponding to $2\times10^{-7}$ in an explicit filter and acting on the highest 33\% of wavenumbers.

\subsection{Cases of study}\label{sec:study_case}

Table \ref{tab:cases} summarizes the studied cases. $Ro_r=(\Omega r)/U_\infty$ is the local radial rotation number, $\lambda=(\Omega R)/V_\infty$ is the tip-speed ratio, $u_{x_\infty}$ and $u_{y_\infty}$ are the resultant, undisturbed inflow velocities in the $x$ and $y$ directions, $Ro_c$ is the rotation number based on the chord, which is equivalent to $\Omega$, and $\Omega_x$ and $\Omega_y$ are the rotation speeds in the $x$ and $y$ directions.

\begin{table}[htb!]
    \centering
    \resizebox{\columnwidth}{!}{
    \begin{tabular}{c c c c c c c c c c} \hline
         $AoA$ ($^\circ$) & $Ro_r$ & $\lambda$ & $V_\infty$ & $\Omega r$ & $u_{x_\infty}$ & $u_{y_\infty}$ & $\Omega$ ($\equiv Ro_c$) & $\Omega_x$ & $\Omega_y$\\ \hline
        12.8 & 3.1 & 4.6  & 0.30 & 0.95  & 0.9751 & 0.2219 & 0.0523 & 0.0044 & 0.0521\\
        12.8 & 0.0 & 0.0  & 0.30 & 0.95  & 0.9751 & 0.2219 & 0      & 0      & 0\\\hline
        4.2  & 6.3 & 9.3  & 0.16 & 0.98  & 0.9973 & 0.0737 & 0.0542 & 0.0045 & 0.0540\\
        4.2  & 0.0 & 0.0  & 0.16 & 0.98  & 0.9973 & 0.0737 & 0      & 0      & 0\\\hline
        1.2  & 9.4 & 13.9 & 0.11 & 0.99  & 0.9998 & 0.0217 & 0.0546 & 0.0046 & 0.0544\\
        1.2  & 0.0 & 0.0  & 0.11 & 0.99  & 0.9998 & 0.0217 & 0      & 0 & 0\\ \hline
    \end{tabular}
    }
    \caption{Parameters of the studied cases.}
    \label{tab:cases}
\end{table}

\subsection{Mesh resolution and spatial/temporal convergence}\label{sec:mesh_spat_conv}

The mesh was generated with the software ICEM \citep{ICEM}, aiming at an implicit large eddy simulation (LES) resolution with a well-resolved wall region. The near-wall grid resolution was assessed using the parameter $\Delta \xi^+_{wall}=\Delta \xi_{wall}/(\nu \sqrt{\tau_w})$, where $\xi = x, y, z$. $\nu$ is the kinematic viscosity, $\tau_w$ is the wall stress. For $\xi = y$, $\Delta y_{wall}$ is taken as the height of the first cell. Considering all cases and both suction and pressure sides, $\Delta y^+_{wall} < 0.4$ for almost the entirety of the domain, except very close to the trailing edge, where it slightly increases to 0.5. For $\xi=x,z$, since the GLL points are not uniformly spaced, $\Delta \xi_{wall}$ is taken as the maximum over each element. The average and minimum $\Delta \xi^+_{wall}$ are related to the maximum by factors of 0.68 and 0.31, respectively. $\max_{elem.} \Delta x^+_{wall} < 23$ for the whole airfoil, and $\max_{elem.} \Delta z^+_{wall} < 10$, except in the first 4\% of the chord, where $\max_{elem.} \Delta z^+_{wall}$ can increase to 15. Note that this is close to the optimum mesh resolution for LES to match DNS, which is  $\Delta y^+_{wall}=0.64$, $\Delta x^+_{wall}=18$,  $\Delta z^+_{wall}=9$, as found by \citet{negi2018a} and as per simulations of turbulent boundary layers by \citet{schlatter2010b}. The inflation rate of the elements in the normal direction to the airfoil and streamwise direction in the wake is 1.15. 

\begin{figure}
    \subfigure[Reynolds stresses on the suction side for 7th and 8th polynomial orders.
    \label{fig:spatial_convergence}]
    {\includegraphics[width=0.495\linewidth,trim={0cm 0cm 0cm 0cm},clip]
        {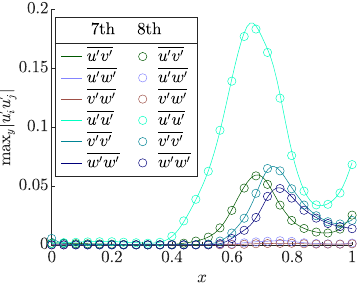}}%
    \subfigure[Pressure coefficient on both sides.
    \label{fig:Cp_temporal_convergence}]
    {\includegraphics[width=0.495\linewidth,trim={0cm 0cm 0cm 0cm},clip]
        {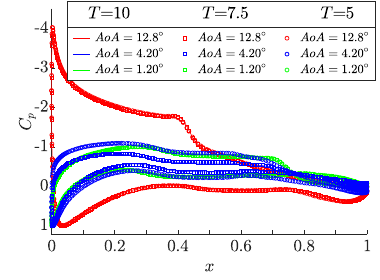}}        
    \caption{Panel (a): spatial convergence of the Reynolds stresses on the suction side for standard (7th order) and refined (8th order) meshes for $AoA=1.2^\circ$, $Ro_r=0$. Panel (b): Temporal convergence of the mean pressure distribution of the rotating cases.}
    \label{fig:grid_resolution}    
\end{figure}

Furthermore, a grid independence study was performed by simulating the $AoA=1.2^\circ$, $Ro_r=0$ case with polynomial order height instead of seven, yielding 179 million grid points. Figure \ref{fig:spatial_convergence} compares the Reynolds stresses in these two meshes. The results are spatially converged for polynomial order seven with a maximum difference of 1.5\% compared to the finer grid. Therefore, the order-seven mesh is selected for all computations in this work. The temporal convergence of mean fields is also assessed concerning the length of the time series. The mean pressure coefficient ($C_p$) averaged over a time of 5, 7.5, and 10 flowthroughs ($T=5$, 7.5, and 10) are compared in Fig. \ref{fig:Cp_temporal_convergence} considering the rotating simulations, i.e., $Ro_r\neq0$. For $AoA=12.8^\circ$ and $AoA=4.2^\circ$, the results are converged for $T=7.5$, although the differences are already minor for $T=5$ in the former case. For $AoA=1.2^\circ$, convergence occurs for $T=5$. The analysis of the skin friction coefficient ($C_f$) led to similar conclusions. Therefore, the selected averaging time of $T=10$ is enough for the convergence of the mean fields for all cases. Each simulation required $3.5 \times 10^{5}$ CPU hours for $T=10$.

%-------------------------------------------------------------------------------
%------------------------------------------------------------------------------

\section{Results} \label{sec:results}

Unless otherwise indicated, the following analyses focus on the suction side of the airfoil, which is more susceptible to separation, transition, and rotation effects.

\subsection{General flow characteristics}
\label{sec:gen_flow_characteristics}

\subsubsection{Mean flow}\label{sec:mean_flow}

\begin{figure}
    \subfigure[Pressure coefficient on both sides.
    \label{fig:Cp_all_cases}]
    {\includegraphics[width=0.495\linewidth,trim={0cm 0cm 0cm 0cm},clip]
        {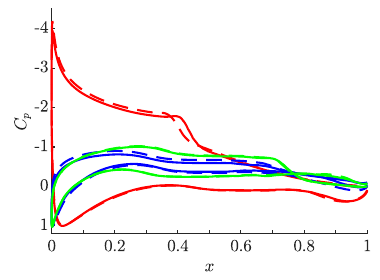}}%
    \subfigure[Friction coefficient on the suction side.
    \label{fig:Cf_all_cases_SS}]
    {\includegraphics[width=0.5\textwidth,trim={0cm 0cm 0cm 0cm},clip]
        {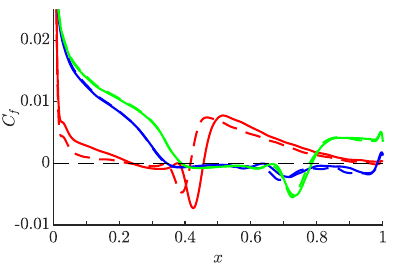}}
    \caption{Effect of rotation on the mean pressure and skin friction distributions. The following color code is used: \protect\redline for $AoA=12.8^\circ$, $Ro_r=3.1$;  \protect\blueline for $AoA=4.2^\circ$, $Ro_r=6.3$;  \protect\greenline for $AoA=1.2^\circ$, $Ro_r=9.4$. Dashed lines correspond to their $Ro_r=0$ counterparts.}
    \label{fig:Cp_Cf}    
\end{figure}

The mean pressure coefficient ($C_p$) is displayed in Fig. \ref{fig:Cp_all_cases}. The $AoA=12.8^\circ$ case presents an adverse pressure gradient (APG) almost on the entirety of the suction side (upper part of the curves) for $AoA=12.8^\circ$, with a peak suction of about $-4$ near the leading edge, characteristic of wind turbine airfoils under similar $AoA$ \citep{ronsten1992}. Nevertheless, the suction peak is 5.6\% more negative for the non-rotating case, a difference that persists up to the location of the maximum LSB height for the $Ro_r=0$  case, which occurs at $x=0.36$. At this point, the non-rotating case presents a sharp pressure increase, associated with laminar-turbulent transition. The pressure rise occurs at $x=0.39$ for the rotating case, indicating a more downstream transition. This suggests that the Coriolis force reduces the APG on the suction side, as also noticed by \citet{dumitrescu2004} and \citet{herraez2014}. However, rotation increases the APG vis-à-vis the non-rotating case in the region extending from $x=0.42$, close to the LSB maximum height, to $x=0.5$, 4\% downstream of reattachment. This phenomenon can be attributed to the momentum deficit in this region, which promotes a reversal in the direction of the Coriolis force that passes to act as an APG. After turbulent reattachment, there are no significant differences between the $C_p$ curves on the suction side. Considering the $AoA=4.2^\circ$ case, the shape of the pressure distribution does not have a marked leading-edge suction peak as observed for $AoA=12.8^\circ$. The region under favorable pressure gradient (FPG) on the suction side extends until $x=0.2$ for the non-rotating case and $x=0.21$ for the rotating one. The APG is mild downstream of that, indicating the presence of an extended region of flow separation. The non-rotating case displays a lower $C_p$ until the transition region, around $70\%$ of the chord length, indicating a similar trend to the $AoA=12.8^\circ$ case. Finally, considering the $AoA=1.2^\circ$ case, an FPG acts until $x=0.27$ on the suction, followed by a low APG leading the flow to reattach around 80\% chord. The pressure is slightly lower for the non-rotating case, consistent with the trend noticed for higher $AoA$. On the pressure side, the non-rotating cases display a slightly higher $C_p$, but this difference is only significant in the $AoA=4.2^\circ$ case. This suggests that rotation has an opposite effect on each side of the airfoil, a consequence of the streamwise velocity being higher than its free-stream value in the accelerating region of the suction side and lower than it in the decelerating region of the pressure side. Therefore, the leading-edge region with strong acceleration is especially susceptible to this effect.

%-------------------------------------------------------------------------------
%------------------------------------------------------------------------------

The friction coefficient ($C_f$) on the suction is presented in Fig. \ref{fig:Cf_all_cases_SS}. The $C_f$ is higher in the rotating case for $AoA=12.8^\circ$ upstream of separation, which occurs at $x=0.23$ for both rotating and non-rotating counterparts. The more elevated $C_f$ indicates that the flow is accelerated in the streamwise direction by rotation in this region. The near-zero negative $C_f$ indicates that separation is mild until $x=0.34$ for $Ro_r=0$ and $x=0.38$ for $Ro_r=3.1$. However, downstream of those locations, a stronger reverse flow occurs, noticed by a sharp drop in $C_f$, which is larger for the rotating case. Although taking place more downstream, the reverse flow is more intense in the non-rotating case, likely due to the reverse Coriolis force that acts as an APG in this part of the LSB. This region is highly unstable, and transition occurs, which leads the flow to reattach at $x=0.46$ and $x=0.42$ for the rotating and non-rotating cases, respectively, followed by a large increase in $C_f$ characteristic of turbulent flow. In regard to the $AoA=4.2^\circ$ case, separation occurs at $x=0.34$ for both rotating and non-rotating cases, which is mild until around 56\% chord. In the region upstream of this point, the $C_f$ curves for both rotating and non-rotating conditions agree. However, downstream of $x=0.56$, a stronger reverse flow is present, which occurs earlier for the non-rotating case, displaying a peak negative $C_f$ at $x=0.69$, whereas this occurs at $x=0.72$ for the rotating case. The flow remains separated until close to the trailing edge of the airfoil, with the non-rotating case exhibiting more negative $C_f$ in most of this region. Similar trends are noticed in the $AoA=1.2^\circ$ case, in which separation occurs at $x=0.38$ but remains weak until around 65\% chord. Upstream of this location, the $C_f$ distributions for $Ro_r=0$ and $Ro_r=9.4$ present minimal differences. Stronger separation occurs slightly upstream for the non-rotating case, with peak negative $C_f$ at $x=0.625$, being 1\% of the chord more downstream for the rotating condition. Transition occurs in this region and leads the flow to reattach at $x=0.79$ for the rotating case and $x=0.78$ for the non-rotating one, suggesting a slightly faster breakdown to turbulence in the former case.

\begin{figure}
    \subfigure[$AoA=12.8^\circ$, $Ro_r=3.1$.
    \label{fig:U_mean_L1_o045t_PP_SS_2}]
    {\includegraphics[width=0.495\linewidth,trim={0cm 0cm 0cm 0cm},clip]
        {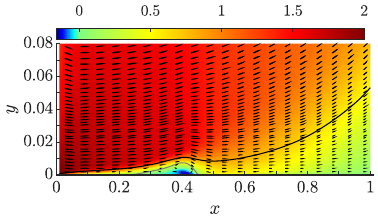}}%
    \subfigure[$AoA=12.8^\circ$, $Ro_r=0$.
    \label{fig:U_mean_L1_o045n_PP_SS_2}]
    {\includegraphics[width=0.495\linewidth,trim={0cm 0cm 0cm 0cm},clip]
        {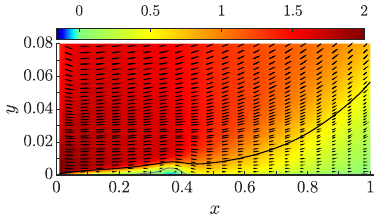}}

    \subfigure[$AoA=4.2^\circ$, $Ro_r=6.3$.
    \label{fig:U_mean_L1_o09t_PP_SS_2}]
    {\includegraphics[width=0.495\linewidth,trim={0cm 0cm 0cm 0cm},clip]
        {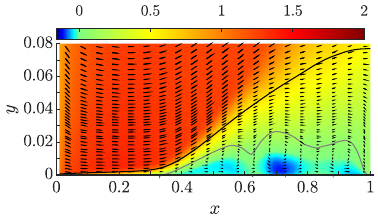}}%
    \subfigure[$AoA=4.2^\circ$, $Ro_r=0$.
    \label{fig:U_mean_L1_o09n2_PP_SS_2}]
    {\includegraphics[width=0.495\linewidth,trim={0cm 0cm 0cm 0cm},clip]
        {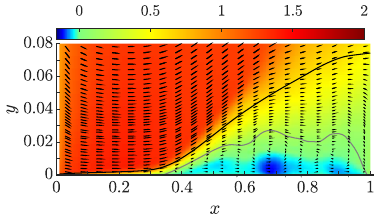}}
        
    \subfigure[$AoA=1.2^\circ$, $Ro_r=9.4$.
    \label{fig:U_mean_L1_o135t_PP_SS_2}]
    {\includegraphics[width=0.495\linewidth,trim={0cm 0cm 0cm 0cm},clip]
        {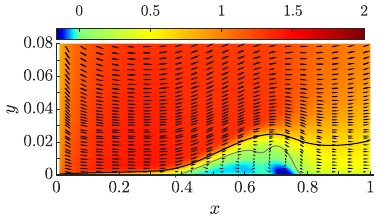}}%
    \subfigure[$AoA=1.2^\circ$, $Ro_r=0$.
    \label{fig:U_mean_L1_o135n_PP_SS_2}]
    {\includegraphics[width=0.495\linewidth,trim={0cm 0cm 0cm 0cm},clip]
        {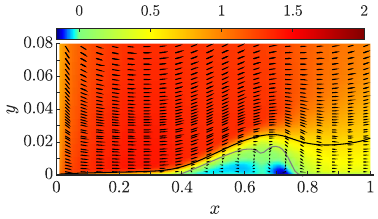}}
    \caption{Spanwise and time-averaged streamwise velocity ($\langle U\rangle_{z, t}$) contours. \protect\blackline is the displacement thickness ($\delta^*$). \protect\grayline is the edge of the LSB ($(x,y) \in \int_{0}^{y} \langle U\rangle_{z, t}(x,\xi) \; d\xi=0$, where $\langle\cdot\rangle_{z, t}$ is the average in span and time). Arrows represent the velocity field.}
    \label{fig:U_mean}    
\end{figure}

The spanwise and time-averaged streamwise velocity field is shown in Fig. \ref{fig:U_mean}, with the displacement thickness ($\delta^*$) depicted with a black line and the edge of the LSB with a gray line. In the $AoA=12.8^\circ$ case, one can notice a stronger reverse flow in the rotating case that reaches -16\% of the free-stream velocity at $x=0.41$ compared to -7\% for the non-rotating case at $x=0.38$ (the LSB is 74.4\% higher in the former case). The reverse flow in the non-rotating case may be enough to trigger absolute instability \citep{alam2000,rodriguez2021}. These locations are slightly downstream of the maximum LSB height ($x=0.4$ and $x=0.36$, respectively). The most intense reverse flow in the rotating case is due to the streamwise deceleration promoted by the Coriolis force accentuated when the flow velocity is substantially lower than its free-stream value, such as inside the LSB. Considering the $AoA=4.2^\circ$ case, the LSB consists of three reverse flow cells located at $x=0.34-0.63$, $x=0.63-0.81$, and $x=0.81-0.99$, with the center cell presenting the strongest reverse flow of -12\% in both cases. Nonetheless, the location of this maximum is more downstream in the rotating case ($x=0.71$) than the non-rotating one ($x=0.68$). These locations correspond approximately to the position of the maximum LSB height. The split of the reverse flow in separate cells is associated with transition to turbulence and inception of unsteadiness \citep{theofilis2000,cherubini2010}. The indentations in the LSB edge are caused by the smaller reverse flow in the contact regions between two adjacent clockwise-rotating recirculation cells, where the flow is mostly oriented in the normal direction. Notice that the LSB height in the rotating case is slightly shorter (-2.57\%); albeit a small reduction, this denotes a different trend from that observed for $AoA=12.8^\circ$. The large $\delta^*$ and LSB size for $AoA=4.2^\circ$ is possibly related to weaker instabilities developing on the separated shear layer, which are not strong enough to make the flow reattach within a short streamwise extent. This probably results from a weak APG. Regarding the $AoA=1.2^\circ$ case, separation occurs at $x=0.38$, and two recirculation vortices are formed at $x=0.38-0.66$ and $x=0.66-0.78 (0.79)$. Transition takes place in the first cell, and reattachment occurs at the end of the second one, located at $x=0.78$ in the non-rotating case and $x=0.79$ in the rotating one. The maximum reverse flow is -16\% for $Ro_r=0$ at $x=0.72$ and -15\% for $Ro_r=9.4$ at $x=0.71$. The LSB is only 1.7\% taller in the rotating case, indicating that the rotation effects seem to be less strong for lower $AoA$, despite the higher rotation speed.

%-------------------------------------------------------------------------------
%------------------------------------------------------------------------------

\begin{figure}
    \subfigure[$AoA=12.8^\circ$, $Ro_r=3.1$.
    \label{fig:W_mean_L1_o045t_PP_SS_2}]
    {\includegraphics[width=0.495\linewidth,trim={0cm 0cm 0cm 0cm},clip]
        {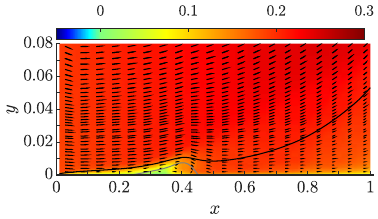}}%
    \subfigure[$AoA=4.2^\circ$, $Ro_r=6.3$.
    \label{fig:W_mean_L1_o09t_PP_SS_2}]
    {\includegraphics[width=0.495\linewidth,trim={0cm 0cm 0cm 0cm},clip]
        {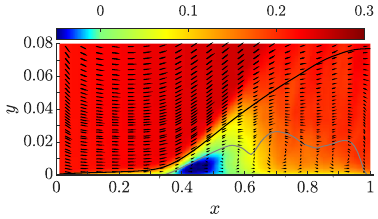}}
        
    \subfigure[$AoA=1.2^\circ$, $Ro_r=9.4$.
    \label{fig:W_mean_L1_o135t_PP_SS_2}]
    {\includegraphics[width=0.495\linewidth,trim={0cm 0cm 0cm 0cm},clip]
        {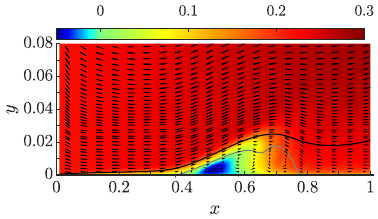}}%
    \caption{Spanwise and time-averaged spanwise velocity ($\langle W\rangle_{z, t}$) contours. \protect\blackline is the displacement thickness ($\delta^*$). \protect\grayline is the edge of the LSB ($(x,y) \in \int_{0}^{y} \langle U\rangle_{z, t}(x,\xi) \; d\xi=0$, where $\langle\cdot\rangle_{z, t}$ is the average in span and time). Arrows represent the velocity field.}
    \label{fig:W_mean}    
\end{figure}

The mean spanwise velocity field is displayed in Fig. \ref{fig:W_mean} for the rotating cases. Results for the non-rotating blade are not presented, as the mean spanwise velocity is zero. In the $AoA=12.8^\circ$ case, this variable is positive on the whole suction side, indicating flow towards the blade root. However, its magnitude is significantly reduced, particularly in the front region of the LSB. The low streamwise velocity in this region reduces the strength of the Coriolis force, pointing towards the blade root, compared to the centrifugal force. Therefore, the spanwise flow is significantly reduced but not reversed since the Coriolis force is still dominant. Also, notice that the spanwise velocity component is significantly low for $y\le\delta^*$ in the laminar region upstream of the LSB and from the separation point to near the location of the maximum LSB height. In the turbulent flow region downstream of the latter location, the low spanwise velocity region is restricted to an area very close to the wall. The $AoA=4.2^\circ$ case displays a region with negative spanwise velocity, i.e., pointing towards the blade tip in the region extending from the separation point to slightly upstream of the maximum height of the first recirculation cell of the LSB. The maximum tip flow reaches 6\% at $x=0.45$, which may be enough to trigger crossflow instability, especially because the spanwise velocity profile is inflectional. The region with spanwise velocity deficit occurs for $y\le\delta^*$ and extends until the end of the first recirculation cell, approximately at $x=0.63$. For $AoA=1.2^\circ$, a tip flow with a maximum intensity of 5\% occurs at $x=0.5$ corresponding to the front part of the first recirculation cell. In this case, there is also the possibility of crossflow instability due to the inflectional crossflow velocity profile \citep{saric2003}. Similar to the previous cases, the area with a deficit of spanwise velocity extends from the leading edge to the end of the first recirculation cell, which in the current case is approximately at $x=0.66$, for $y\le\delta^*$.

%-------------------------------------------------------------------------------
%------------------------------------------------------------------------------

\begin{figure}[ht]
	\centering
	\includegraphics[width=1\linewidth,trim={0cm 0cm 0cm 0cm},clip]{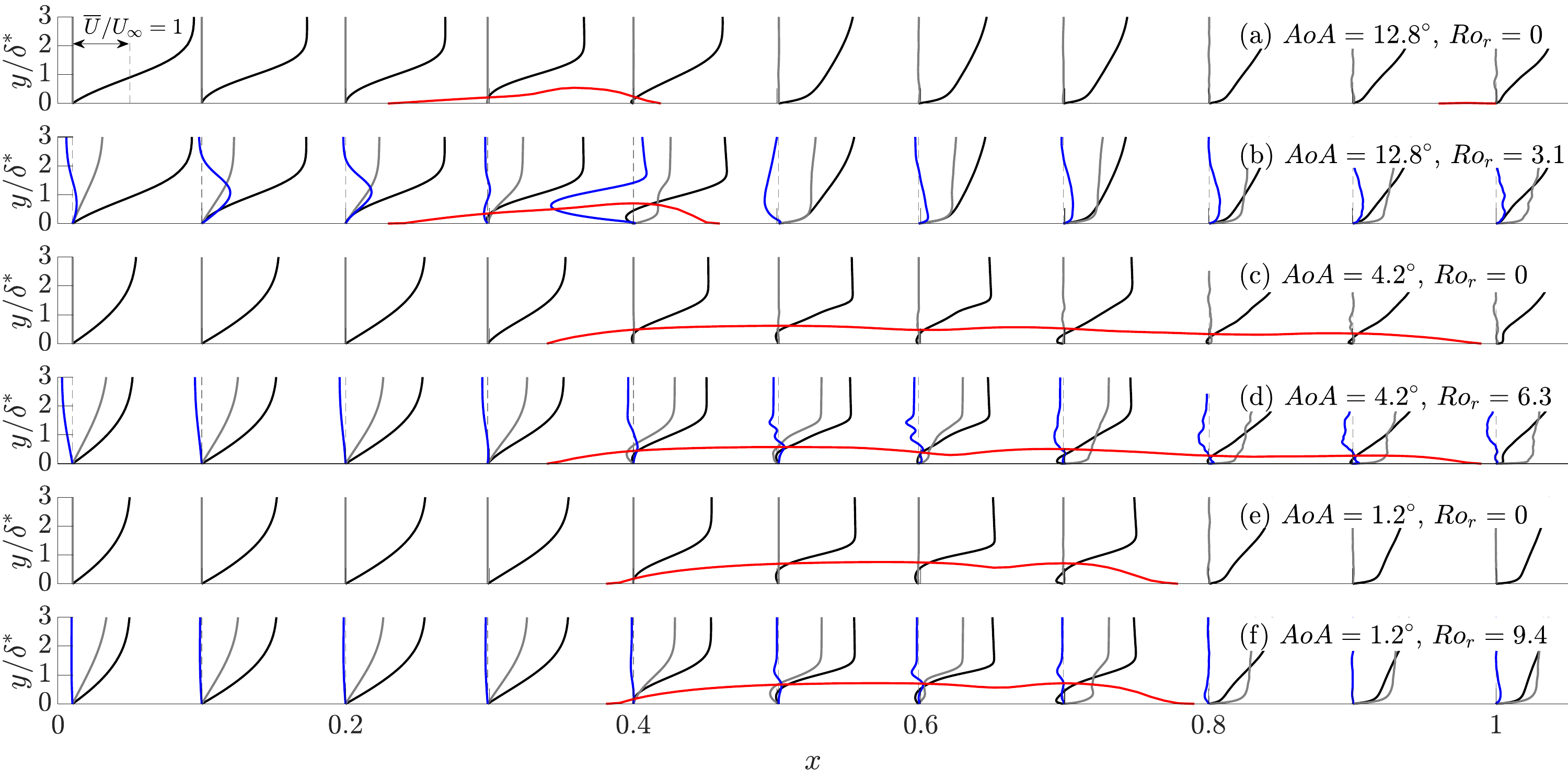}
	\caption{Spanwise and time-averaged profiles of streamwise \protect\blackline ($\langle U \rangle_{z,t}$) and spanwise \protect\grayline ($\langle W \rangle_{z,t}$) velocity. \protect\blueline is the difference between the streamwise velocity in the rotating and non-rotating cases. \protect\redline is the edge of the LSB ($(x,y) \in \int_{0}^{y} \langle U\rangle_{z, t}(x,\xi) \; d\xi=0$, where $\langle\cdot\rangle_{z, t}$ is the average in span and time). The scale of \protect\blackline ($\overline{U}=\langle U \rangle_{z,t}$) is shown in the upper left corner. \protect\grayline and \protect\blueline are plotted with a scale three times larger for enhanced visibility.}
	\label{fig:mean_vel_prof}
\end{figure}

Figure \ref{fig:mean_vel_prof} shows the normal profiles of mean streamwise and spanwise velocity (black and gray solid lines, respectively). Moreover, the difference between the streamwise velocity profiles in the rotating and non-rotating cases ($\overline{\Delta U}$) is depicted with a blue line. Notice that the scale of the gray and blue lines is enlarged three times compared to that of the streamwise velocity. Considering the $AoA=12.8^\circ$, $Ro_r=3.1$ case, the flow is streamwise accelerated compared to the non-rotating case upstream of the LSB, highlighted by the positive values of $\overline{\Delta U}$. The peak in this variable occurs at $y=\delta^*$, the region most affected by rotation. However, the fact that $\overline{\Delta U}\rightarrow0$ in the free-stream indicates that the streamwise velocity in this region is only mildly affected by rotation. Interestingly, inside the LSB, particularly at $x=0.4$, there is a strong deceleration of the flow in the rotating case, demonstrated by the $\overline{\Delta U}<0$ profile at this station. This effect tends to vanish downstream of the LSB, and the flow is accelerated by rotation again for $x\ge0.6$, albeit with a much lower intensity since the turbulent flow presents a smaller near-wall region with a momentum deficit. The spanwise velocity profiles do not possess sign inversion, as previously noticed in the contours of this variable. Nevertheless, the profiles develop a distinct shape inside the LSB, first with a near-zero spanwise velocity at $x=0.3$ and then with a clearly inflectional character at $x=0.4$. Regarding the $AoA=4.2^\circ$, $Ro_r=6.3$ case, the flow is slightly retarded upstream of separation and mildly accelerated near the wall in the LSB. This constitutes a change in the trend with respect to $AoA=12.8^\circ$. This may be connected with the fuller streamwise velocity profiles in the $AoA=4.2^\circ$ case due to the FPG acting for $x<0.21$, which reduces the near-wall momentum deficit and makes the rotation effects accelerate the flow downstream in this region preceding separation. At some stations inside the LSB, the flow is near-stagnant, such as at $x=0.4$, allowing a mild acceleration of the flow by the Coriolis force. The spanwise velocity profiles are clearly inflectional for some locations inside the LSB, particularly for $x=0.4$, $0.5$, and $0.6$, with the first two stations displaying tip flow ($\overline{W}<0$). Regarding the $AoA=1.2^\circ$ case, the effects of rotation on the streamwise velocity are practically zero outside the LSB and mild inside it. The spanwise velocity profiles indicate near-wall tip flow at $x=0.5$ and clear inflectional character at $x=0.5$, $0.6$, and $0.7$.

%------------------------------------------------------------------------------
%------------------------------------------------------------------------------

Table \ref{tab:flowcharSS} summarizes some characteristics of the LSB and mean flow. $x_{b_{1}}$ and $x_{b_{2}}$ are the leading and trailing edges of the separation region. $x_{y_{max}}$ and $y_{max}$ are the streamwise location of the separation bubble maximum height, and the value of this height ($y_{max}/\delta^*$ for the height normalized by the displacement thickness). $x_a$ is the position of the start of the APG. $u_{r_{max}}$ and $w_{r_{max}}$ are the maximum streamwise and spanwise negative flows, respectively. $x_{u_{r_{max}}}$ and $x_{w_{r_{max}}}$, in this order, are the streamwise locations where these maxima occur.

\begin{table}[ht]
\centering
\caption{Characteristics of the mean flow.}
\label{tab:flowcharSS}
\resizebox{\columnwidth}{!}{
\begin{tabular}{c c c c c c c c c c c c}
& & \\ % put some space after the caption
\hline
\hline
AoA ($^\circ$) & $Ro_r$ & $x_{b_{1}}$ & $x_{b_{2}}$ & $x_{y_{max}}$ & $y_{max}\times10^2$ & $y_{max}/\delta^*$ & $x_{a}$ & $u_{r_{max}}$ & $x_{u_{r_{max}}}$ & $w_{r_{max}}$ & $x_{w_{r_{max}}}$ \\
\hline
12.8 & 3.1  & 0.23	& 0.46 & 0.40 & 0.75 & 0.70 & 0.00 & -0.16 & 0.41 & 0 .00 & -    \\
12.8 & 0.0  & 0.23	& 0.42 & 0.36 & 0.43 & 0.54 & 0.00 & -0.07 & 0.38 & -     & -    \\\hline
4.2  & 6.3  & 0.34	& 0.99 & 0.70 & 2.65 & 0.51 & 0.21 & -0.12 & 0.71 & -0.06 & 0.45 \\
4.2  & 0.0  & 0.34	& 0.99 & 0.68 & 2.72 & 0.56 & 0.20 & -0.12 & 0.68 & -     & -    \\\hline
1.2  & 9.4 & 0.38  & 0.79 & 0.70 & 1.78 & 0.71 & 0.27 & -0.15 & 0.71 & -0.05 & 0.50 \\
1.2  & 0.0  & 0.38  & 0.78 & 0.70 & 1.75 & 0.71 & 0.27 & -0.16 & 0.72 & -     & -    \\
\hline
\hline
\end{tabular}
}
\end{table}

\begin{figure}
    \subfigure[max$_y \overline{u'v'}$.
    \label{fig:uv_SS}]
    {\includegraphics[width=0.495\linewidth]
        {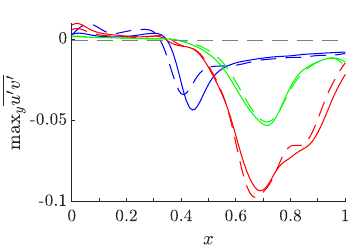}}%
    \subfigure[max$_y \overline{v'w'}$.
    \label{fig:vw_SS}]
    {\includegraphics[width=0.495\linewidth]
        {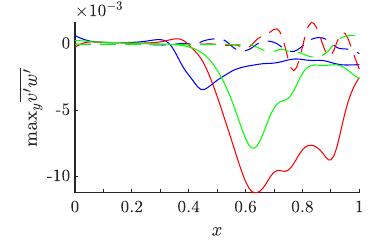}}
        
    \subfigure[max$_y u'_{RMS}$.
    \label{fig:urms_SS}]
    {\includegraphics[width=0.495\linewidth]
        {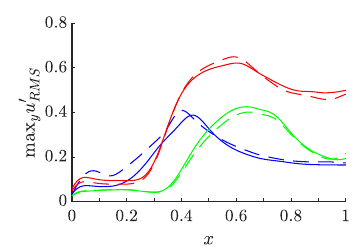}}%
    \subfigure[max$_y w'_{RMS}$.
    \label{fig:wrms_SS}]
    {\includegraphics[width=0.495\linewidth]
        {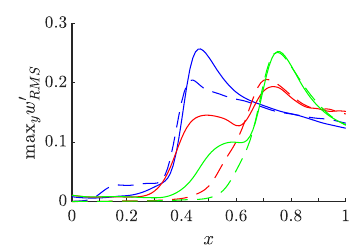}}
    \caption{Spanwise averaged Reynolds stresses and root mean square (RMS) values of streamwise, normal, and spanwise velocity perturbations. The following color code is used: \protect\blueline for $AoA=12.8^\circ$, $Ro_r=3.1$;  \protect\redline for $AoA=4.2^\circ$, $Ro_r=6.3$;  \protect\greenline for $AoA=1.2^\circ$, $Ro_r=9.4$. Dashed lines correspond to their $Ro_r=0$ counterparts.  \protect\dashedgrayline is the $\overline{u' v'}=-0.001$ line.}
    \label{fig:reynolds_stresses}    
\end{figure}

Figure \ref{fig:reynolds_stresses} shows the Reynolds stresses and the root mean square (RMS) values of velocity perturbations as a function of the streamwise coordinate. $\overline{u' v'}$ (Fig. \ref{fig:uv_SS}) becomes negative as instabilities develop in the flow. \citet{yuan2005} proposed the threshold $\overline{u' v'}=-0.001$ to identify transition to turbulence, which tends to pinpoint the early stages of transition instead of the full turbulent breakdown. For $AoA=12.8^\circ$, this criterion is met at $x=0.33$ in the non-rotating case and $x=0.35$ in the rotating one, indicating a downstream shift of the transition location by rotation. The streamwise location of the maximum boundary layer shape factor $H=\delta^*/\theta$, where $\theta$ is the momentum thickness, is also an indicator of the transition location \citep{burgmann2008}. These locations are  $x=0.36$ for $Ro_r=0$ and $x=0.39$ for $Ro_r=3.1$. Therefore, transition is delayed by 2\%-3\% by rotation in the $AoA=12.8^\circ$ case. $\overline{u' v'}$ reaches a more negative peak value in the rotating case, suggesting a stronger breakdown to turbulence. Regarding $AoA=4.2^\circ$, the $\overline{u' v'}=-0.001$ threshold is reached earlier in the rotating case ($x=0.35$) than in the non-rotating one ($x=0.37$), suggesting a slight upstream shift of transition with rotation. However, the maximum $H$ occurs at $x=0.5$ in both cases, indicating that, despite a faster beginning of transition in the rotating case, this process is completed at the same location as in the non-rotating case. Furthermore, the peak in $\overline{u' v'}$ is more pronounced in the non-rotating case this time, indicating a stronger breakdown for this case. The rotation effects on the $AoA=1.2^\circ$ case are qualitatively similar to those in the $AoA=4.2^\circ$ case, although less pronounced. Yuan's criterion is met 1\% more upstream in the rotating case ($x=0.35$), pointing to an earlier transition start. Nonetheless, the full turbulent breakdown occurs at the same place, yielding a transition point at $x=0.7$ in both cases. The rotating cases also present significantly higher $\overline{v' w'}$ component, as presented in Fig. \ref{fig:vw_SS}, indicating the formation of coherent structures, e.g., streamwise vortices. This is particularly true for cases with large separation regions, such as $AoA=4.2^\circ$ and $AoA=1.2^\circ$. Stall cells have been observed experimentally and numerically in non-rotating \citep{winkelmann1980,bippes1982,schewe2001} and rotating \citep{yang2006,diottavio2008,raghav2014} wings. \citet{rodriguez2010} indicated that such cells in a non-rotating stalled airfoil could be associated with a steady three-dimensional global mode, as found by \citet{theofilis2000}. Another possibility is that coherent structures arise from crossflow instability \citep{gross2012}. 

Figures \ref{fig:urms_SS} and \ref{fig:wrms_SS} present the RMS values of streamwise ($u'$) and spanwise ($w'$) velocity perturbations. In the $AoA=12.8^\circ$ case, all disturbance components are higher for $Ro_r=0$ upstream of $x=0.43$, indicating that the boundary layer is more unstable in at least part of this region in the non-rotating case. Downstream of this point, $w'_{RMS}$ develop higher peak values in the rotating case, whereas the peak in $u'_{RMS}$ is comparable to those in the non-rotating case. This suggests the presence of structures in the $yz$ plane for $Ro_r=3.1$ that could be either crossflow modes or non-linearly/non-modal generated streamwise vortices \citep{schmid2001}. A threshold of $u'_{RMS}\approx40\%$ of the free-stream velocity precedes transition, even though this value is smaller compared to the local edge velocity: 25.8\% at $x=0.4$ for $Ro_r=0$ and 25.9\% at $x=0.44$ for $Ro_r=3.1$. Considering $AoA=4.2^\circ$ and $AoA=1.2^\circ$, the RMS values of all perturbations are higher in the rotating case for a large portion of the suction side for $AoA=4.2^\circ$ and in all this region for $AoA=1.2^\circ$. In particular, $w'_{RMS}$ displays considerably higher values in the rotating case until $x\approx0.65$. At $x=0.46$, for example, $w'_{RMS}$ is 947\% higher in the rotating case compared to $Ro_r=0$ for $AoA=4.2^\circ$  and 1316\% for $AoA=1.2^\circ$. Relative to the local boundary-layer edge velocity, $w'_{RMS}$ corresponds to 11.3\% and 3.9\% for $AoA=4.2^\circ$ and $AoA=1.2^\circ$, respectively. Notice that $x=0.46$ is located before transition, suggesting that these perturbations do not arise from non-linear effects but rather from a possible crossflow instability. The higher values of $u'_{RMS}$ for $AoA=1.2^\circ$ in the rotating case could also indicate the increase in the amplification of pre-existing instability mechanisms, e.g., TS and KH modes.

%-------------------------------------------------------------------------------
%------------------------------------------------------------------------------

% ok
In order to identify the inflectional character of the flow and the role of inviscid instabilities, the ratio ($\mathcal{R}$) between the normal locations of the inflection point ($y_{in}$) in the tangential ($U_{t}$) and crossflow ($U_{cr}$) velocity profiles \citep{saric2003}, and the locus of maximum production of perturbation kinetic energy ($y_{max_{\mathcal{P}}}$) is analyzed. $\mathcal{R} \rightarrow 1$ indicates the dominance of an inviscid instability mechanism as opposed to a viscous one when  $\mathcal{R} \rightarrow 0$ \citep{diwan2009}. The tangential velocity profiles are inflectional on the whole suction side for $AoA=12.8^\circ$ due to the strong APG. However, the non-rotating case reaches $\mathcal{R}=1$ significantly more upstream ($x=0.12$) than the rotating one ($x=0.32$), indicating that a predominantly inviscid instability Tollmien-Schlichting (TS) or Kelvin-Helmholtz (KH) mode develops first in the $Ro_r=0$ case, likely due to the counteraction of the APG by the Coriolis force in this region in the rotating case. In the $AoA=4.2^\circ$ case, $\mathcal{R} = 1$ is reached at $x=0.43$ for $Ro_r=0$ and $x=0.46$ for $Ro_r=6.3$, whereas for $AoA=1.2^\circ$ this occurs at $x=0.53$ for $Ro_r=0$ and $x=0.54$ for $Ro_r=9.4$, indicating reversed trend to that for higher $AoA$. These results indicate that an inviscid instability of the tangential velocity profiles plays an important role in all cases, and rotation considerably shifts downstream this instability for $AoA=12.8^\circ$ and, to a lesser extent, $AoA=4.2^\circ$ and moves it slightly upstream for $AoA=1.2^\circ$. The crossflow velocity profiles in the rotating cases become inflectional and thus susceptible to crossflow instability since the leading edge. However, $\mathcal{R} \not\rightarrow 1$ indicates that if a crossflow instability exists, it is not the dominant mechanism in the flow.

Table \ref{tab:flowtrans} outlines the streamwise locations where the mean streamwise ($x_{i_U}$), tangential ($x_{i_{U_t}}$), spanwise ($x_{i_W}$), and crossflow ($x_{i_{U_{cr}}}$) velocity profiles become inflectional. The transition locations are also summarized, based on the streamwise location of the  $-\overline{u v}/U_\infty^2=0.001$ threshold \citep{yuan2005} ($x_{tr_1}$)  and the maximum boundary-layer aspect ratio ($x_{tr_2}$)  \citep{burgmann2008}.

\begin{table}[ht]
\centering
\caption{Characteristics of the inflectional velocity profiles and transition locations.}
\label{tab:flowtrans}
\begin{tabular}{c c c c c c c c}
& & \\ % put some space after the caption
\hline
\hline
AoA ($^\circ$) & $Ro_r$ & $x_{i_U}$ & $x_{i_{U_t}}$ & $x_{i_W}$ & $x_{i_{U_{cr}}}$ & $x_{tr_1}$ & $x_{tr_2}$ \\
\hline
12.8 & 3.1 & 0.00 & 0.00 & 0.00 & 0.00 & 0.35 & 0.39\\
12.8 & 0.0 & 0.00 & 0.00 & -    & -    & 0.33 & 0.36\\\hline
4.2  & 6.3 & 0.20 & 0.22 & 0.07 & 0.00 & 0.35 & 0.50\\
4.2  & 0.0 & 0.21 & 0.22 & -    & -    & 0.36 & 0.50\\\hline
1.2  & 9.4 & 0.26 & 0.29 & 0.08 & 0.00 & 0.40 & 0.59\\
1.2  & 0.0 & 0.27 & 0.29 & -    & -    & 0.41 & 0.59\\
\hline
\hline
\end{tabular}
\end{table}

\subsubsection{Instantaneous flow structures}\label{sec:instat_flow}

\begin{figure}
    \subfigure[$AoA=12.8^\circ$, $Ro_r=3.1$.
    \label{fig:uprime2_L1_o045t}]
    {\includegraphics[width=0.495\linewidth,trim={0cm 3.1cm 0cm 6cm},clip]
        {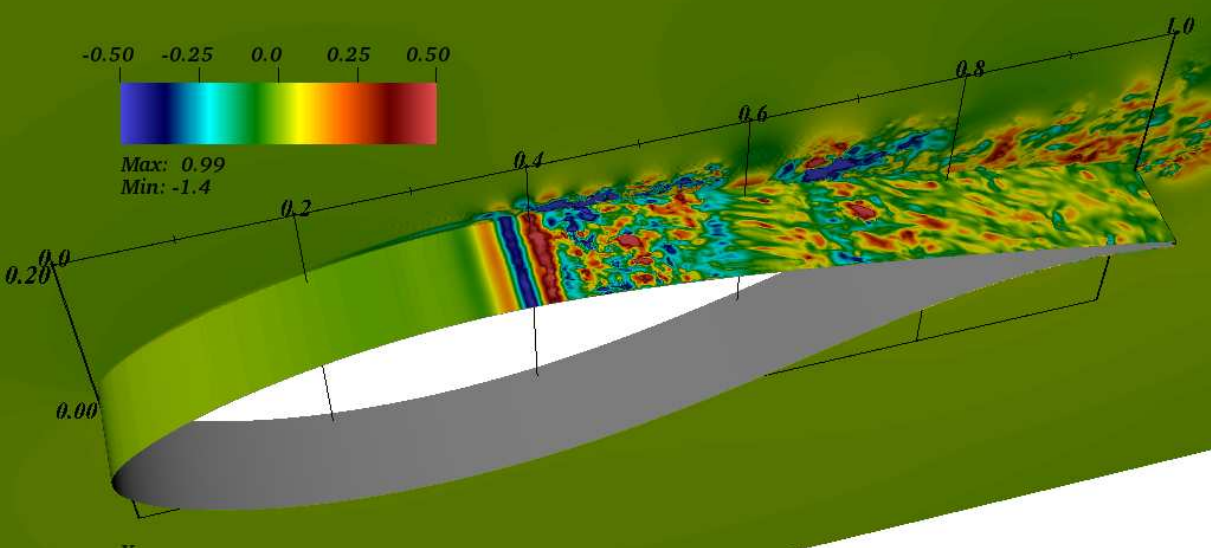}}%
    \subfigure[$AoA=12.8^\circ$, $Ro_r=0$.
    \label{fig:uprime2_L1_o045n}]
    {\includegraphics[width=0.495\linewidth,trim={0cm 3.1cm 0cm 6cm},clip]
        {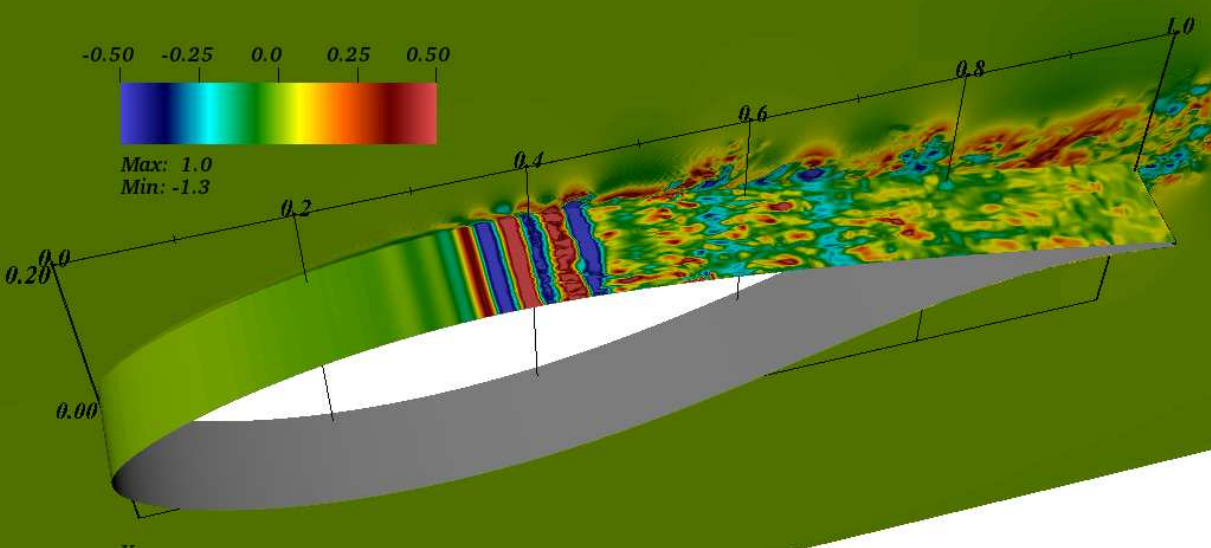}}

    \subfigure[$AoA=4.2^\circ$, $Ro_r=6.3$.
    \label{fig:uprime2_L1_o09t}]
    {\includegraphics[width=0.495\linewidth,trim={0cm 3.1cm 0cm 6cm},clip]
        {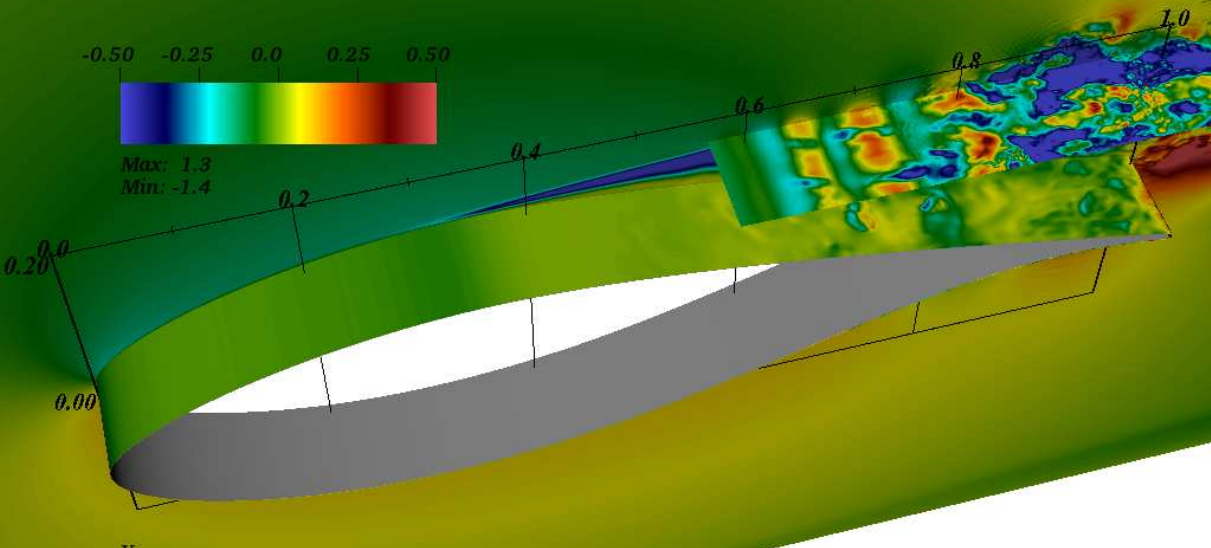}}%
    \subfigure[$AoA=4.2^\circ$, $Ro_r=0$.
    \label{fig:uprime2_L1_o09n2}]
    {\includegraphics[width=0.495\linewidth,trim={0cm 3.1cm 0cm 6cm},clip]
        {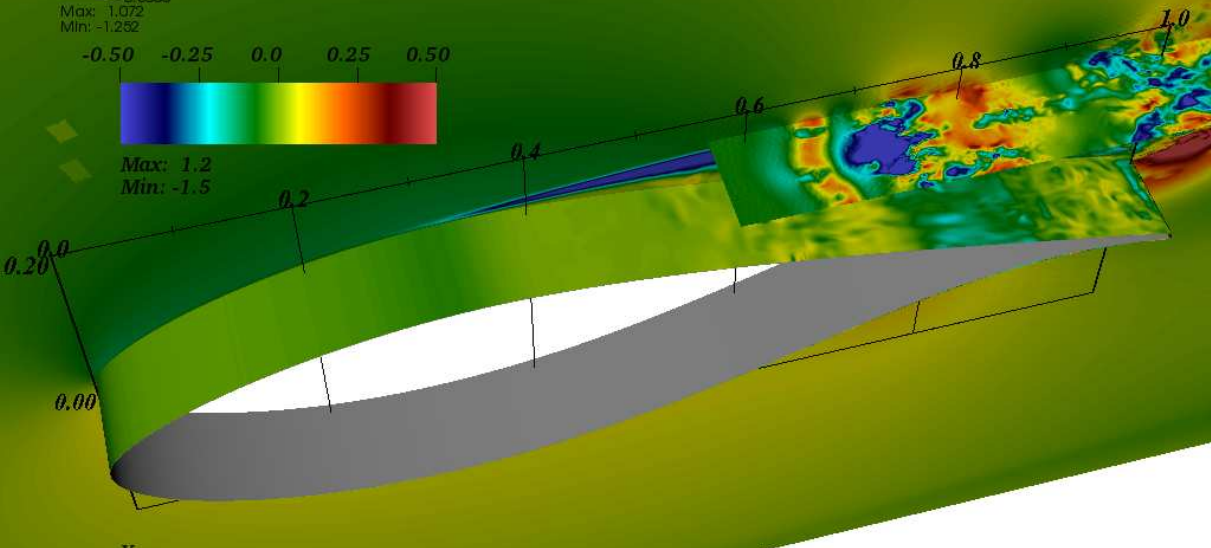}}
        
    \subfigure[$AoA=1.2^\circ$, $Ro_r=9.4$.
    \label{fig:uprime2_L1_o135t}]
    {\includegraphics[width=0.495\linewidth,trim={0cm 3.1cm 0cm 6cm},clip]
        {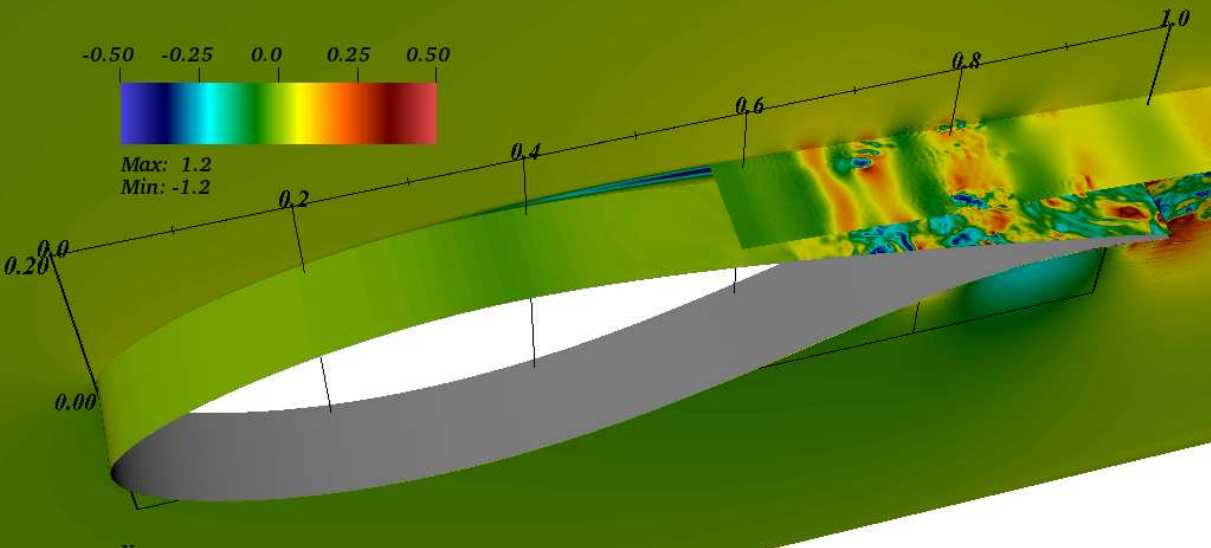}}%
    \subfigure[$AoA=1.2^\circ$, $Ro_r=0$.
    \label{fig:uprime2_L1_o135n}]
    {\includegraphics[width=0.495\linewidth,trim={0cm 3.1cm 0cm 6cm},clip]
        {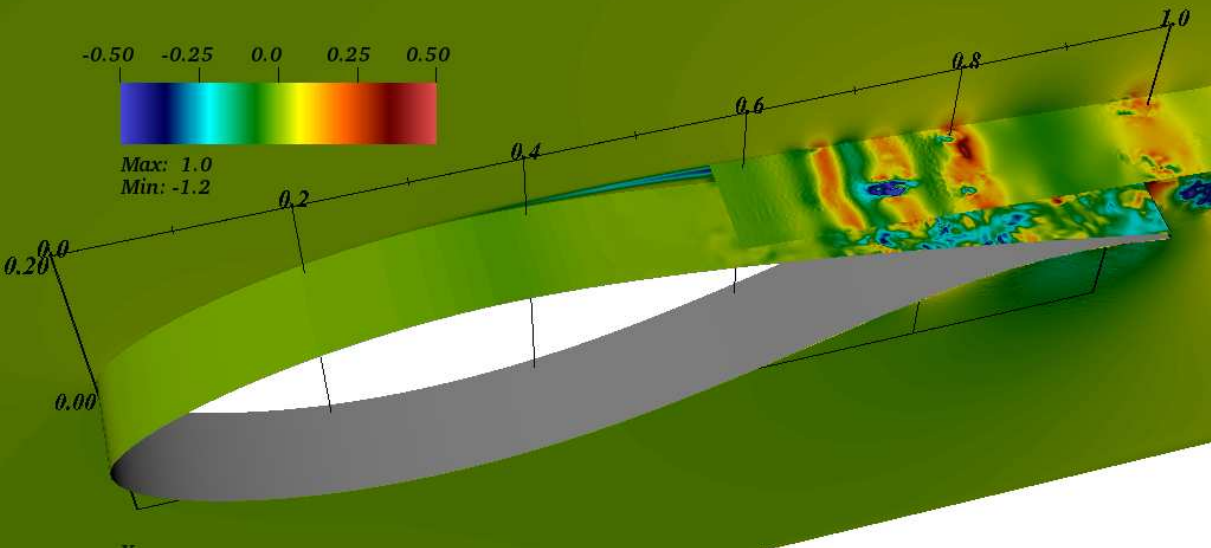}}
    \caption{Contours of streamwise velocity fluctuations ($u'=U-\langle U \rangle_{z,t}$) near the wall, on a longitudinal plane, and, for $AoA=4.2^\circ$ and $AoA=1.2^\circ$, on a plane parallel to the separated shear layer.}
    \label{fig:uprime}    
\end{figure}

In order to have a deeper insight into the instantaneous flow structures, Fig. \ref{fig:uprime} presents the contours of streamwise velocity fluctuations ($u'$) over a near-wall and a longitudinal plane. In the $AoA=4.2^\circ$ and $AoA=1.2^\circ$ cases, a plane passing through the shear layer is also considered. For $AoA=12.8^\circ$, spanwise-uniform, streamwise oscillating structures form downstream of $x\approx0.35$ in the rotating case, 5\% upstream of the LSB maximum height. These rolls likely correspond to KH rolls due to the shear layer roll-up \citep{brinkerhoff2011}. Notice that the most unstable mode smoothly transition from a TS instability in the attached shear layer to a KH mode, which occurs close to the LSB maximum height \citep{diwan2009}. The fluctuation amplitude at $x=0.4$ is 36\% of $U_\infty$ or 21\% of the local edge velocity. Transition and breakdown into small-scale turbulence occur at this location, even though rolls seem to persist further downstream, as shown in the longitudinal plane and also observed by \citet{hosseinverdi2019}. Streamwise streaks appear downstream of $x=0.58$, already in the turbulent region. The abrupt breakdown and the streaks are characteristic of an oblique secondary instability mechanism \citep{rist_vki}. However, this requires further analysis. Notice that the streaks are tilted in the positive $z$ direction, indicating that rotation promotes asymmetry regarding positive and negative spanwise wavenumbers. Regarding the non-rotating case, spanwise-uniform structures appear shortly downstream of separation, at $x\approx0.25$, much more upstream than in the rotating case. However, spanwise coherent rolls can be noticed until a more downstream location ($x\approx0.45$) compared to the rotating case, suggesting a slower breakdown. Furthermore, streaks perfectly aligned with the streamwise direction form shortly downstream of this breakdown. This may indicate a different secondary instability mechanism compared to the rotating case or an enhancement of the same mechanism in the latter case.

Considering $AoA=4.2^\circ$, featuring a long separation bubble, the primary instabilities develop mainly on the edge of the shear layer. Over a plane at this location, rolls are formed downstream of $x=0.6$, probably corresponding to a KH-dominated instability due to the significant distance from the wall. According to \citet{jaroslawski2023}, the increased distance of the shear layer from the wall could foster three-dimensional global instabilities such as the mechanism found by \citet{rist2002} and \citet{rodriguez2021}. In the non-rotating case, these structures are spanwise modulated with a wavelength corresponding to the domain width. In contrast, in the rotating case, the rolls seem to remain two-dimensional prior to the formation of turbulent spots. Finally, the $AoA=1.2^\circ$ case displays a wavepacket forming downstream of $x=0.6$ for rotating and non-rotating conditions, also corresponding to KH rolls. These structures display a spanwise modulation preceding their breakdown, with the mechanism seemingly the same.

%-------------------------------------------------------------------------------
%------------------------------------------------------------------------------

\subsection{Flow stability and perturbation evolution}
\label{sec:pert_evol_flow_stab}

\subsubsection{Stability analyses}\label{sec:stab}

Primary local stability of the mean flow profiles was performed by solving the spatial stability problem of the linearized Navier-Stokes (LNS) equations with rotation terms, given by

\begin{equation}\label{eq:stab_prob}
        \mathbf{\mathcal{A}} \mathbf{\hat{q}} = \alpha \mathbf{\mathcal{B}} \mathbf{\hat{q}},
\end{equation}

\begin{equation}
\mathbf{\mathcal{A}} = \left( \begin{array}{cccc}
0   & \mathcal{D} & i \beta & 0\\
\Xi & -\mathcal{D} \overline{\mathbf{U}} & -2 \Omega_y \overline{\mathbf{W}} & 0\\
0   & \Xi & 2 \Omega_x & -\mathcal{D}\\
2 \Omega_y & -\mathcal{D} \overline{\mathbf{W}} - 2 \Omega_x & -i \beta & 0 \\
\end{array} \right), \quad
\mathbf{\mathcal{B}} = \left( \begin{array}{cccc}
-i & 0 & 0 & 0 \\
i \overline{\mathbf{U}} & 0 & 0 & i \\
0 & i \overline{\mathbf{U}} & 0 & 0 \\
0 & 0 & i \overline{\mathbf{U}} & 0 \\
\end{array} \right),
\end{equation}

\noindent where $\Xi=1/Re_c(\mathcal{D}-\beta^2)-i(\beta \overline{\mathbf{W}} - \omega)$, $\mathcal{D}=\partial/\partial y$, $i=\sqrt{-1}$, $\omega = 2 \pi f$, $Re_c=U_\infty c/\nu$, and $\mathbf{q'}(x,y,z,t)=[u' v' w' p']^{T} = \mathbf{\hat{q}}(y,\beta,\omega) \exp{\left[i(-\omega t + \alpha x + \beta z)\right]}$ is the vector of velocity and pressure perturbations. Notice that the assumption of small streamwise variations allowed dropping high order $\alpha$ terms, reducing the cost of solving the eigenvalue problem for ample parametric space. The boundary conditions were $u',v',w',p'=0$ at $y=0$ and $y\rightarrow\infty$. The solution of the stability problem \ref{eq:stab_prob} involved the discretization of the $y$ coordinate with 150 Chebyshev points, which was proven to yield converged results.

\begin{figure}
    \subfigure[Rotating versus non-rotating cases.
    \label{fig:max_alphai_f_beta_x}]
    {\includegraphics[width=0.495\linewidth]
        {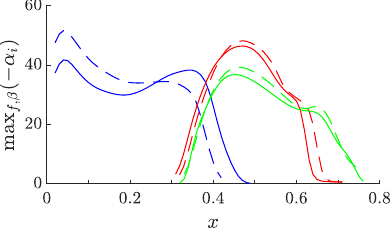}}%
    \subfigure[Rotating cases with and without $W$.
    \label{fig:max_alphai_f_beta_x_W0}]
    {\includegraphics[width=0.495\linewidth]
        {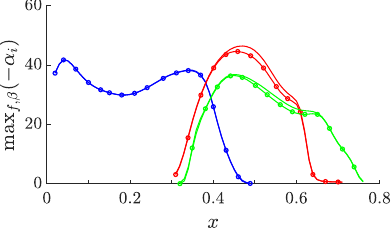}}
    \caption{Local growth rate from spatial stability analysis. The following color code is used: \protect\blueline for $AoA=12.8^\circ$, $Ro_r=3.1$;  \protect\redline for $AoA=4.2^\circ$, $Ro_r=6.3$;  \protect\greenline for $AoA=1.2^\circ$, $Ro_r=9.4$. Dashed lines correspond to their $Ro_r=0$ counterparts. Circle lines denote rotating cases ($Ro_r\neq0$) with $W=0$ in the stability analysis.}
    \label{fig:max_alphai}    
\end{figure}

The local growth rates ($-\alpha_i$) for rotating and non-rotating cases are compared in Fig. \ref{fig:max_alphai_f_beta_x}. The $AoA=12.8^\circ$ non-rotating case presents higher growth rates upstream of $x=0.26$, a point located 3\% downstream of the separation location. This is likely due to the stabilization of the boundary layer upstream by the Coriolis force that reduces the APG. However, rotation passes to decelerate the flow inside the LSB, destabilizing the flow and leading to higher growth rates. Notice that the small region where the growth rates of the primary instability in the rotating case are higher than in the non-rotating case is not enough to counteract the stabilization of the pre-separation flow. Therefore, transition is more downstream in the rotating case. For $AoA=4.2^\circ$, the first unstable mode appears at $x=0.31$, slightly upstream of the separation location ($x=0.34$). From $x=0.31$ to $x=0.42$, corresponding to the first recirculation cell, the growth rates are slightly higher in the rotating case, a trend that is reversed further downstream due to the streamwise deceleration of the flow by rotation. Notice that the regions where rotation destabilizes the flow reverse for the $AoA=12.8^\circ$ and $AoA=4.2^\circ$ cases, being the separated region in the former and the pre-separation one in the latter. The $AoA=1.2^\circ$ case follows the same trend as the  $AoA=4.2^\circ$ one, in which $-\alpha_i$ of the rotating case is higher than the non-rotating one upstream of separation ($x=0.38$) and the opposite in the LSB.

In order to distinguish if amplification changes by rotation are related to the modification of the streamwise or spanwise velocity, the stability analysis of the rotating cases neglecting the spanwise velocity is performed. The results presented in Fig. \ref{fig:max_alphai_f_beta_x_W0} indicate that the effect of the spanwise velocity component is negligible in the $AoA=12.8^\circ$ case. However, the spanwise velocity generated by rotation destabilizes the flow in the $AoA=4.2^\circ$  and $AoA=1.2^\circ$ cases in regions corresponding to negative near-wall spanwise velocity (tip flow), which are highly inflectional (Fig. \ref{fig:W_mean}). This suggests that crossflow modes render the flow slightly more unstable in these cases, but a deeper analysis is required to assess the nature of the instability. Notice that, since $\overline{W}$ is stabilizing, the higher $-\alpha_i$ observed in the non-rotating cases in certain regions (Fig. \ref{fig:max_alphai_f_beta_x}) must be related to more unstable streamwise velocity profiles.

%-------------------------------------------------------------------------------
%------------------------------------------------------------------------------

\begin{figure}
    \subfigure[TS/KH mode.
    \label{fig:KH_mode_LNS_L1_045t_f15_b10_x033}]
    {\includegraphics[width=0.495\linewidth]
        {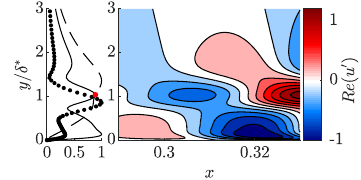}}%
    \subfigure[Crossflow mode.
    \label{fig:crossflow_mode_LNS_L1_09t_f0_b188.5_x046}]
    {\includegraphics[width=0.495\linewidth]
        {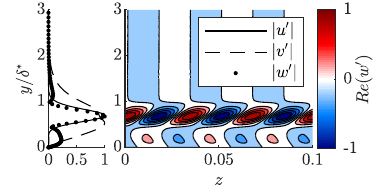}}
    \caption{Perturbation profiles and contours of (a) TS/KH mode for $AoA{=}12.8^\circ$, $Ro_r{=}3.1$, $f{=}15$, $\beta{=}10$, $x{=}0.33$, and (b) crossflow mode for $AoA{=}4.2^\circ$, $Ro_r{=}6.3$, $f{=}0$, $\beta{=}188.5$, $x{=}0.46$ from local spatial stability analysis. The red dot on the profile of the left panel indicates the normal location of the inflection point in the mean streamwise velocity.}
    \label{fig:LNS_profiles}    
\end{figure}

In the $AoA=12.8^\circ$ case, the maximum amplification occurs for $f=15$, $\beta=10$ in the rotating condition, whereas the corresponding locus is $f=12$, $\beta=0$ in the non-rotating condition. The shift from $\beta=0$ to $\beta=10$ results from the spanwise flow in the $+z$ direction in the rotating case. Notice that $\beta=0$ perturbations are the most unstable in two-dimensional flows, like the non-rotating case \citep{schmid2001}. The rotation terms in the stability equations have limited effects on the results. The modes responsible for this instability are of mixed nature containing both the influence of TS waves near the wall and KH instability close to the inflection point location. The relative importance of each mechanism varies between cases and with the streamwise location, but essentially the further away the shear layer lies from the wall, the more preponderant the KH mechanism \citep{diwan2009}, such as in the $AoA=4.2^\circ$ case. The most unstable mode will be denominated TS/KH instability. The most crucial role of rotation is to generate a near-wall peak in the spanwise velocity perturbation ($w'$) of the TS/KH mechanism. The normal perturbation profiles of the mode with maximum amplification for $AoA=12.8^\circ$, $Ro_r=3.1$ are presented in Fig. \ref{fig:KH_mode_LNS_L1_045t_f15_b10_x033}. The $u'$ profile displays a near-wall peak corresponding to the TS mechanism. This is followed by a phase jump and a second amplitude maximum at the location of the inflection point of the mean streamwise velocity profile (red dot), characteristic of a KH instability \citep{rist_vki}.  It is also important to analyze the propagation angle $\Psi=\arctan(\beta/\alpha_r)-\arctan(\overline{W}_e/\overline{U}_e)$, i.e., the angle between the wavevector ($\alpha_r,\beta$) and the inviscid streamline ($\overline{U}_e,\overline{W}_e$) in the $xz$ plane. For $Ro_r=3.1$, $\Psi =[-98.6^\circ,81.1^\circ]$ and for $Ro_r=0$, $\Psi=[- 85.1^\circ, 85.1^\circ]$. Essentially, $\Psi<0$ is mostly located on the $\beta<0$ semi-plane and vice-versa. An exception occurs when modes display $\alpha_r<0$, which may allow $\Psi<0$ modes on the $\beta>0$ semi-plane. Notice that $\Psi$ reaches more negative values for the rotating case. \citet{borodulin2019} found traveling crossflow modes with $\Psi=[84^\circ,87^\circ]$ or $\Psi=[-97^\circ,-92^\circ]$ in a swept wing, whereas stationary crossflow modes were clustered at $\Psi\pm89^\circ$. Therefore, rotation may enable crossflow modes, but the growth rates were low compared to the TS/KH instability. The reason for that may be the not very inflectional spanwise velocity profiles, similar to the rotating disk with axial inflow \citep{hussain2011,deschamps2017}.

\begin{figure}
    \subfigure[Local growth rate.
    \label{fig:alphai_f_beta_L1_o09t_x046}]
    {\includegraphics[width=0.495\linewidth]
        {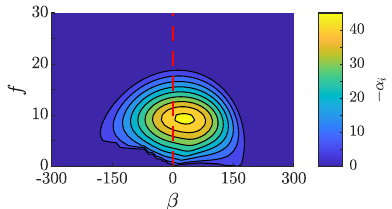}}%
    \subfigure[Propagation angle.
    \label{fig:psi_x046_L1_09t_4_6_merged}]
    {\includegraphics[width=0.495\linewidth]
        {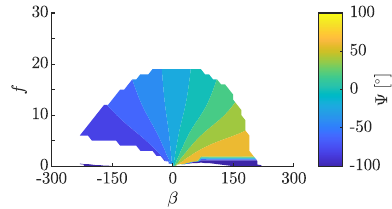}}
    \caption{Contours of local growth rate ($-\alpha_i$) and propagation angle ($\Psi$) from local spatial stability analysis for $AoA=4.2^\circ$, $Ro_r=6.3$, at $x=0.46$.
    }
    \label{fig:alphai_f_beta}    
\end{figure}

For $AoA=4.2^\circ$, Fig. \ref{fig:alphai_f_beta_L1_o09t_x046} indicates that the highest $-\alpha_i$ occurs at $f=9$ and $\beta=30$ at $x=0.46$ in the rotating condition, corresponding to a TS/KH instability. The propagation angle is $\Psi=[-100.4^\circ,77.2^\circ]$, as shown in Fig. \ref{fig:psi_x046_L1_09t_4_6_merged}, where this variable is shown for unstable modes only. Low-frequency $\beta\neq0$ modes become unstable in the rotating case, including stationary disturbances, which is more visible for $\beta>0$, although there is a small unstable region close to the $f=0$ axis for $\beta<0$. These modes present high $|\Psi|$ but low growth rates and may be crossflow modes. The profiles and contours of the stationary mode for $\beta=188.5$ ($\Psi=-95.5^\circ$, $-\alpha_i=3.7$) are presented in Fig. \ref{fig:crossflow_mode_LNS_L1_09t_f0_b188.5_x046}. The eigenfunction and contours of spanwise velocity disturbances agree with those from the literature for a crossflow mode \citep{saric2003,borodulin2019}. In the non-rotating case, the maximum amplification occurs at $f=10$ and $\beta=0$. The low-frequency unstable region with $\beta\neq0$ is absent in this case, and $\Psi=[-83.7^\circ,83.7^\circ]$, with much smaller $|\Psi|$ than the rotating case. This supports the appearance of crossflow modes due to rotation. At last, the $AoA=1.2^\circ$ case indicates maximum growth at $f=11$, $\beta=20$ for $Ro_r=9.4$, and $f=11$, $\beta=0$ for $Ro_r=0$, both consisting of a TS/KH mode. Furthermore,  $\Psi=[-101.1^\circ,83.5^\circ]$ in the former and $\Psi=[-84.2^\circ,84.2^\circ]$, also suggesting the appearance of crossflow vortices in the rotating cases. However, unlike the $AoA=4.2^\circ$ case, only non-stationary vortices are present, similar to the $AoA=12.8^\circ$ case. Their growth rates are low compared to the TS/KH mode.

%-------------------------------------------------------------------------------
%------------------------------------------------------------------------------

%-------------------------------------------------------------------------------
%------------------------------------------------------------------------------

\begin{figure}
    \subfigure[$AoA=12.8^\circ$, $Ro_r=3.1$.
    \label{fig:neutral_curve_L1_o045t_2}]
    {\includegraphics[width=0.495\linewidth]
        {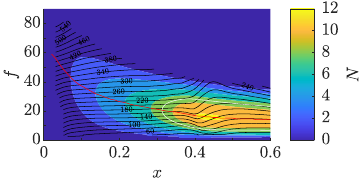}}%
    \subfigure[$AoA=12.8^\circ$, $Ro_r=0$.
    \label{fig:neutral_curve_L1_o045n_3}]
    {\includegraphics[width=0.495\linewidth]
        {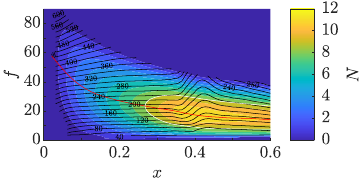}}

    \subfigure[$AoA=4.2^\circ$, $Ro_r=6.3$.
    \label{fig:neutral_curve_L1_o09t_2}]
    {\includegraphics[width=0.495\linewidth]
        {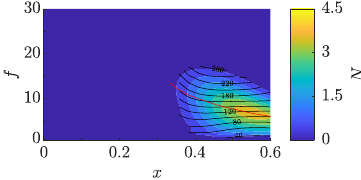}}%
    \subfigure[$AoA=4.2^\circ$, $Ro_r=0$.
    \label{fig:neutral_curve_L1_o09n2_2}]
    {\includegraphics[width=0.495\linewidth]
        {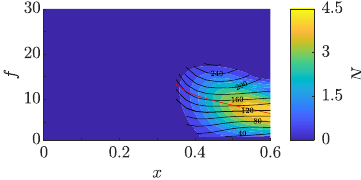}}
        
    \subfigure[$AoA=1.2^\circ$, $Ro_r=9.4$.
    \label{fig:neutral_curve_L1_o135t_2}]
    {\includegraphics[width=0.495\linewidth]
        {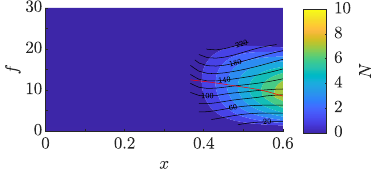}}%
    \subfigure[$AoA=1.2^\circ$, $Ro_r=0$.
    \label{fig:neutral_curve_L1_o135n_2}]
    {\includegraphics[width=0.495\linewidth]
        {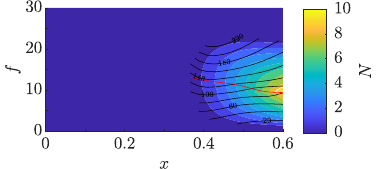}}
    \caption{Contours of $N$ factor obtained with PSE for $\beta=0$ using ($\langle \mathbf{U} \rangle_{z, t}$) as the base flow. \protect\blackline are the isolines of the real part of the streamwise wavenumber ($\alpha_r$), and \protect\redline denotes the loci of maximum amplification. The white line corresponds to $N=9$ (threshold for natural transition).
    }
    \label{fig:neutral_curve}    
\end{figure}

The $N$ factor is a more relevant parameter to predict transition \citep{arnal2000}. For clean environments, $N=9$ is often assumed as a reliable threshold for natural transition. This parameter was computed with the parabolized stability equations (PSE) code NOLOT \citep{hanifi1994}. Here the $N$ factor is based on the total kinetic energy of the perturbations. The results for $\beta=0$ are presented in Fig. \ref{fig:neutral_curve}, as a function of the frequency and streamwise coordinate. For $AoA=12.8^\circ$, the maximum $N$ factor is $N=12.1$ for $f=16$ at $x=0.43$ in the rotating case and $N=11.6$ for $f=17$ at $x=0.4$ in the non-rotating one. At those locations, the streamwise wavelength $\lambda_x=2\pi/\alpha_r$ is 0.05, agreeing with that of the rolls in Fig. \ref{fig:uprime2_L1_o045n}. The results also indicate the stabilization of high-frequency TS waves close to the leading edge. For $AoA=4.2^\circ$, the rotating case reaches $N=3.8$ for  $f=7$ at $x=0.54$, whereas the non-rotating case presents $N=4.4$ also for $f=7$ but at $x=0.59$. This low $N$ factor could be due to an early start of secondary instability or the inception of absolute or global instabilities  \citep{huerre1990,theofilis2000},  not captured by the PSE method. Finally, considering $AoA=1.2^\circ$, the maximum $N$ factor is $8.1$ for $f=8.6$ in the rotating case and $N=10.7$ for $f=9.5$ in the non-rotating one, which occurs 1\% downstream of the transition location ($x=0.59$). In summary, the $N$ factor reaches higher maximum values in the rotating case for $AoA=12.8^\circ$ and in the non-rotating case otherwise. Note that the transition location is more upstream (Table \ref{tab:flowtrans}) in the rotating case for $AoA=12.8^\circ$. This is not a contradiction, as the higher $N$ factor in the rotating case only occurs more downstream than the transition location in the non-rotating condition, as rotation stabilizes the boundary layer upstream of the LSB and destabilizes it in the region of strong reverse flow.

%-------------------------------------------------------------------------------
%------------------------------------------------------------------------------

\begin{figure}
    \subfigure[Complex $\alpha$ plane.
    \label{fig:cusp_map_alpha_2}]
    {\includegraphics[width=0.495\linewidth]
        {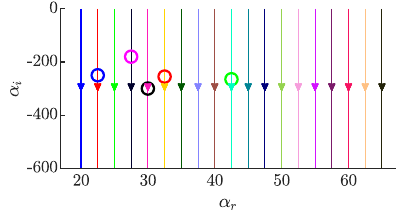}}%
    \subfigure[Growth rate of the absolute instability.
    \label{fig:cusp_map_alphai_x_2}]
    {\includegraphics[width=0.495\linewidth]
        {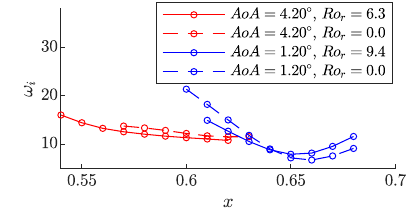}}%

    \subfigure[$\alpha\rightarrow\omega$, $AoA=4.2^\circ$, $Ro_r=6.3$, $x=0.57$.
    \label{fig:cusp_map_omega_L1_o09t_x057_2}]
    {\includegraphics[width=0.495\linewidth]
        {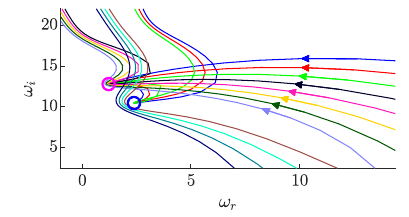}}%
    \subfigure[$\alpha\rightarrow\omega$, $AoA=4.2^\circ$, $Ro_r=0$, $x=0.57$.
    \label{fig:cusp_map_omega_L1_o09n2_x057_2}]
    {\includegraphics[width=0.495\linewidth]
        {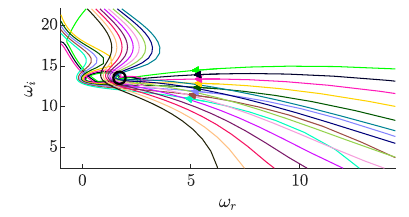}}
        
    \subfigure[$\alpha\rightarrow\omega$, $AoA=1.2^\circ$, $Ro_r=9.4$, $x=0.66$.
    \label{fig:cusp_map_omega_L1_o135t_x066_2}]
    {\includegraphics[width=0.495\linewidth]
        {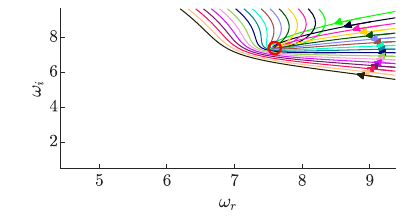}}%
    \subfigure[$\alpha\rightarrow\omega$, $AoA=1.2^\circ$, $Ro_r=0$, $x=0.66$.
    \label{fig:cusp_map_omega_L1_o135n_x066_2}]
    {\includegraphics[width=0.495\linewidth]
        {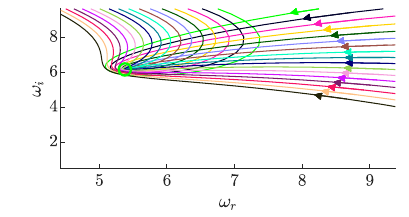}}
    \caption{Complex mapping $\alpha\rightarrow\omega$ via the Orr-Sommerfeld operator. The $\omega$ image for $AoA=4.2^\circ$ (panels (c) and (d)) were obtained at $x=0.57$ and for $AoA=1.2^\circ$ (panels (e) and (f)) at $x=0.66$. The unstable cusp is shown with a circle in both $\omega$ (panels (c), (d), (e), and (f)) and $\alpha$ (panel (a)) planes. The imaginary part of $\omega$, corresponding to the growth rate of the absolute instability, is shown in panel (b).}
    \label{fig:cusp_map}    
\end{figure}

Sufficiently high reverse flows can cause the boundary layer to behave as an oscillator in the so-called absolute instability \citep{huerre1990}. Thresholds for its inception may be reverse flows above 20\% \citep{hammond1998}, $15~\%-20~\%$ \citep{alam2000}, and $6~\%-8~\%$  \citep{rodriguez2013,hosseinverdi2013,rodriguez2021}. Nevertheless, other parameters, such as the Reynolds number and the distance of the inflection point from the wall, also play a role \citep{rist_vki,avanci2019}.
The cusp map \citep{kupfer1987}, i.e., the Orr-Sommerfeld/Squire mapping $\alpha\rightarrow\omega$, where $\alpha\in\mathbb{C}$, was obtained to assess whether convective/absolute instabilities exist. The anstaz for the normal velocity ($v'$) and vorticity ($\eta'$) perturbations is $\mathbf{q'}(x,y,z,t)=[v' \eta']^{T} = \mathbf{\hat{q}}(y,\beta,\omega) \exp{\left[i(-\omega t + \alpha x + \beta z)\right]}$. Thus, a cusp in the $\omega$ trajectories in the $\omega_i>0$ semi-plane indicates an absolute instability. Figure \ref{fig:cusp_map_alpha_2} shows the complex $\alpha$ plane mapped on $\omega$. No unstable cusp exists for $AoA=12.8^\circ$, suggesting that the LSB is convectively stable. For $AoA=4.2^\circ$, two unstable cusps with frequencies $f=1.2$ and $f=2.4$ at $x=0.57$ were found in the rotating case (Fig. \ref{fig:cusp_map_omega_L1_o09t_x057_2}), whereas only one appears in the non-rotating one, at $f=1.7$. The absolutely unstable region for $AoA=4.2^\circ$ is $x=0.54-0.62$ in the first condition (Fig. \ref{fig:cusp_map_alphai_x_2}) and $x=0.57-0.63$ for $Ro_r=0$, with the growth rates being slightly higher in the latter scenario. These LSBs are likely globally unstable as a finite region of absolute instability suggests the occurrence of global instability \citep{huerre1990}. The results for $AoA=1.2^\circ$, at $x=0.66$, indicate an unstable cusp at $f=8$ in the rotating case (Fig. \ref{fig:cusp_map_omega_L1_o135t_x066_2}), and at $f=5.4$ in the non-rotating one. The rotating case region of absolute instability is $x=0.61-0.68$, and $x=0.60-0.68$ for the non-rotating one. The absolute growth rates are lower in the rotating case for $x\le0.64$, possibly due to the streamwise acceleration promoted by rotation. Notice that absolute instability may be related to the split of the LSB in more than one recirculation cell \citep{cherubini2010}.

%-------------------------------------------------------------------------------
%------------------------------------------------------------------------------

\subsubsection{Spectral analysis}\label{sec:spectral}

\begin{figure}
    \subfigure[$AoA=12.8^\circ$, $Ro_r=3.1$.
    \label{fig:sigma_spectrum_f_b_L1_o045t_2}]
    {\includegraphics[width=0.495\linewidth]
        {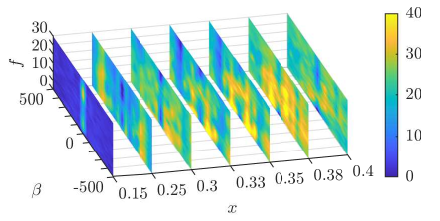}}%
    \subfigure[$AoA=12.8^\circ$, $Ro_r=0$.
    \label{fig:sigma_spectrum_f_b_L1_o045n_2}]
    {\includegraphics[width=0.495\linewidth]
        {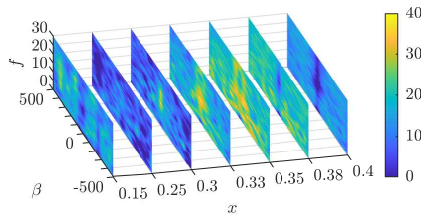}}

      \subfigure[$AoA=4.2^\circ$, $Ro_r=6.3$.
    \label{fig:sigma_spectrum_f_b_L1_o09t_2}]
    {\includegraphics[width=0.495\linewidth]
        {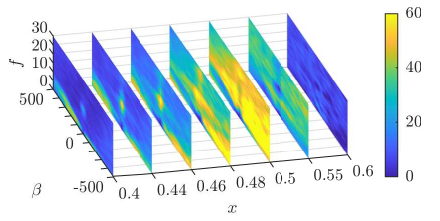}}%
    \subfigure[$AoA=4.2^\circ$, $Ro_r=0$.
    \label{fig:sigma_spectrum_f_b_L1_o09n_2}]
    {\includegraphics[width=0.495\linewidth]
        {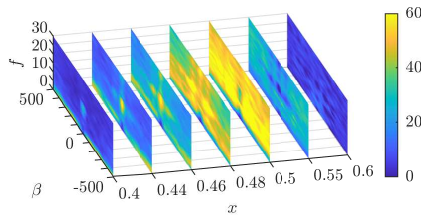}}

    \subfigure[$AoA=1.2^\circ$, $Ro_r=9.4$.
    \label{fig:sigma_spectrum_f_b_L1_o135t_2}]
    {\includegraphics[width=0.495\linewidth]
        {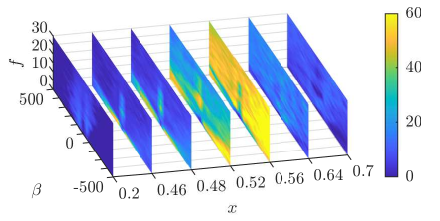}}%
     \subfigure[$AoA=1.2^\circ$, $Ro_r=0$.
    \label{fig:sigma_spectrum_f_b_L1_o135n_2}]
    {\includegraphics[width=0.495\linewidth]
        {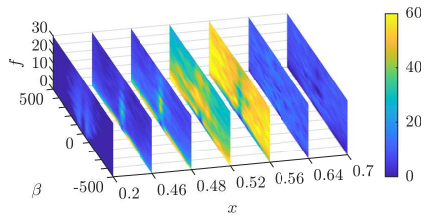}}     
    \caption{Growth rate ($\sigma$) spectra at several streamwise stations.}
    \label{fig:spectra_sigma}    
\end{figure}

%ok
The spectra of the growth rate $\sigma = |u'|^{-1} \frac{\partial |u'|}{\partial x}$ at several streamwise slices are presented in Fig. \ref{fig:spectra_sigma}. Disturbances with frequency $f$ and spanwise wavenumber $\beta$ will be labeled with the pair  $(f/f_f,n_z)$, where $f_f$ is the fundamental instability frequency, to be identified in each case, and $n_z = \beta L_z/(2 \pi)$, where $L_z$ is the domain width. For $AoA=12.8^\circ$, rotating case, TS waves display high $\sigma$ for $f=23$, $\beta=0$ at $x=0.25$. The nature of this instability can be confirmed by linear stability analysis, as done before, or by noting that $F=\frac{2 \pi f \nu 10^6}{\overline{U}_e^2}=663.5$ and $Re_{\delta^*}=\frac{\overline{U}_e \delta^*}{\nu}=444.1$ at $x=0.15$ for $f=23$ is close to the most unstable TS waves in a Falkner-Skan profile under APG \citep{obremski1969}. At $x=0.25$, several modes are excited over the $\beta=0$ axis due to flow separation, and non-linearly generated modes appear for $\beta=\pm 314$ ($n_z=\pm 5$). The mode $(1,0)$ ($f_f=15$), corresponding to the most-unstable TS/KH instability, is also present for $x\ge0.25$ but displays lower $\sigma$ than secondary disturbances. At $x=0.33$ and $x=0.35$, the $(0,\pm 2)$ modes, representing streamwise vortices, present the highest $\sigma$. Moreover, the growth of the $(1,0)$ and $(1,\pm 2)$ modes become apparent at $x=0.35$, where $(1,-2)$ grows more than $(1,2)$, explaining the tilted streaks in the simulation. This confirms the presence of Marxen's oblique mechanism \citep{marxen2003} involving the triad $(1,0)$, $(0,\pm 2)$, and $(1,\pm 2)$. The second-generation modes $(0,\pm 2)$ possibly result from the oblique mechanism involving $(1,\pm1)$ waves \citep{rist_vki}. The lack of symmetry around $\beta=0$ is due to the rotation-generated spanwise flow that renders one side of the spectra more unstable. At $x=0.38$, shortly before transition ($x=0.39$), the amplitude of the $(1,0)$ mode becomes substantial, and the sub-harmonic mode $(0.5,-2)$ develops the highest $\sigma$. In the non-rotating case (Figs. \ref{fig:sigma_spectrum_f_b_L1_o045n_2}), due to more unstable TS waves near the leading edge compared to the rotating case, which triggers early non-linear interactions, the spectrum at $x=0.15$ is already noisy. At $x=0.3$, the TS/KH mode with $f_f=12$ presents high $\sigma$, and, at $x=0.35$, preceding transition ($x=0.36$), the non-linearly generated oblique mode $(1,\pm1)$ shows an important amplification, with the generation of sub-harmonics. Marxen's oblique mechanism involving $(\cdot,\pm 2)$ modes seems not dominant in the non-rotating case.

The results for $AoA=4.2^\circ$ indicate that disturbances with $f \lessapprox 5$ display significant $\sigma$ in the rotating case, especially for $\beta>0$, for $x \le 0.48$. The TS/KH instability appears at $f_f=13.5$, $\beta=0$, for $x \le 0.46$. At $x=0.48$, the maximum $\sigma$ shifts to the $(1,1)$ mode. Linear stability analysis predicts a shift of the most unstable TS/KH mode to $\beta=30$ due to rotation and not to $\beta=62.8$. Therefore,  the $(1,1)$ mode is likely to be non-linearly generated. Transition occurs at $x=0.5$, noticed by the high $\sigma$, especially for low frequencies and $\beta<0$. At this location, $(1,\pm2)$ and $(0.5,\pm2)$ modes are also excited, especially for $\beta<0$. The lack of symmetry around the $\beta=0$ axis is again an effect of rotation. In the non-rotating case, the $f=13.5$, $\beta=0$ mode displays the highest $\sigma$ for $x \le 0.46$, similarly to the rotating case, corresponding to the TS/KH instability. For $0.44 \le x \le 0.48$, two lobes appear corresponding to the $(1,\pm 2)$ modes. Unlike the rotating case, these modes present the highest $\sigma$ and are symmetrical around the $\beta=0$ axis. At $x=0.48$, the sub-harmonics $(0.5,\pm 2)$ also develop important amplification, and transition is triggered. Marxen's oblique mechanism seems only important in the non-rotating case. Lastly, the $AoA=1.2^\circ$ case displays a high $\sigma$ at $f_f=9.6$, $\beta=0$ for $0.44 \le x \le 0.56$, agreeing with the frequency of the most amplified TS/KH instability predicted with linear stability analysis. This mode is present in the rotating and non-rotating cases. In the rotating case, low-frequency, $\beta\neq0$ disturbances present high $\sigma$ in this $x$ range, especially for $\beta>0$, possibly corresponding to crossflow modes. At $x=0.56$, the sub-harmonic mode $(0.5,\pm 2)$ appears in $\sigma$, where the $(0.5,-2)$ is more amplified. This leads to a transition scenario where particularly low-frequency, $\beta<0$ disturbances grow first, noticed in the heterogeneity in $|u'|$ at $x=0.56$ around the $\beta=0$ axis, similarly to $AoA=4.2^\circ$. The non-rotating condition presents higher $\sigma$ for the TS/KH mode. At $x=0.52$, two lobes are formed at $\beta=\pm 125.7$, suggesting the generation of $(1,\pm 2)$ modes, besides the presence of high $\sigma$ for $(0,\pm 2)$, symmetric around the $\beta=0$ axis. This indicates the occurrence of Marxen's mechanism. Sub-harmonic modes $(0.5,\pm2)$ appear at $x=0.56$, triggering transition. Similar to $AoA=4.2^\circ$, the oblique mechanism involving $(\cdot,\pm 2)$ waves appears not dominant in the rotating case.

%-------------------------------------------------------------------------------
%------------------------------------------------------------------------------

\begin{figure}[!htb]
    \centering
    \includegraphics[width=1\linewidth,trim={0cm 0cm 0cm 0cm},clip]{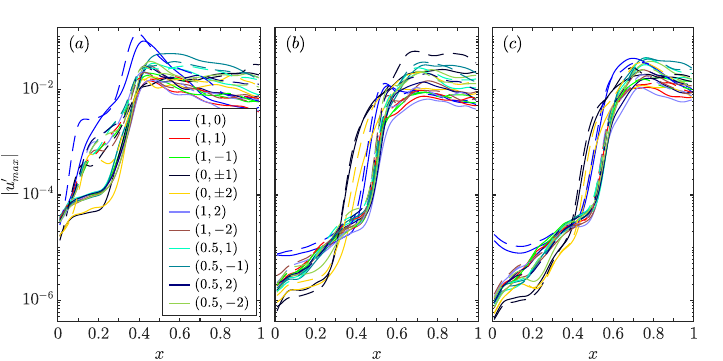}
    \caption{Amplitude spectra from numerical simulation as a function of the streamwise coordinate. Panel (a): $AoA=12.8^\circ$, $f_f=15$ for $Ro_r=3.1$ and $f_f=12$ for $Ro_r=0$. Panel (b): $AoA=4.2^\circ$, $f_f=13.5$. Panel (c): $AoA=1.2^\circ$, $f_f=9.6$. $\beta=0$ in all cases. Solid lines correspond to the rotating case and dashed line to the non-rotating one.}
    \label{fig:sigma_umax_versus_x}
\end{figure}

% ok
The streamwise evolution of $|u'|$ for the main modes is shown in Fig. \ref{fig:sigma_umax_versus_x}. For $AoA=12.8^\circ$, the primary instability, consisting of the TS/KH mode, presents a lower amplitude before transition with a more downstream maximum in the rotating case. Due to the higher growth rates of the $(1,0)$ mode near the leading edge, all secondary modes also display growth in this region. However, in the rotating case, these modes only develop higher amplitudes for $x\ge0.23$, i.e., after separation. The secondary modes with peak amplitude correspond to the sub-harmonic disturbances $(0.5,\pm1)$, followed by the second-generation modes $(0.5,\pm2)$. These modes present higher maximum amplitude in the rotating case, which could explain the faster breakdown of the TS/KH rolls. The more pronounced decay of the $(1,0)$ mode after its peak value indicates a faster loss of spanwise coherence. 

The $AoA=4.2^\circ$ results indicate that the $(0,\pm2)$ modes undergo strong amplification starting at $x=0.25$. These modes could be generated via an oblique mechanism involving $(1,\pm1)$ modes \citep{rist_vki}. The amplitude of the non-rotating disturbances is generally higher considering both primary and secondary modes. The $(0,\pm1)$ modes reach much higher amplitude in the non-rotating and are responsible for the spanwise modulation of the TS/KH rolls, as noticed in the simulations. On the other hand, the sub-harmonic mode $(0.5,-1)$ is the strongest in the non-rotating case. In the $AoA=1.2^\circ$ case, the $(1,0)$ mode displays amplification, particularly downstream of $x=0.4$, with initially higher amplitude for the non-rotating case supplanted by the rotating case downstream of $x=0.5$. The streamwise vortices $(0,\pm1)$ and $(0,\pm2)$ present growth since the leading edge, similar to the $AoA=4.2^\circ$ case, and become the most important modes in the flow in a region prior to the saturation of the primary modes. After that, the sub-harmonic $(0,\pm1)$ and $(0,\pm2)$ modes have the highest amplitude, spanwise modulating the rolls, as seen in the simulations. 

%-------------------------------------------------------------------------------
%-------------------------------------------------------------------------------

Spectral proper orthogonal decomposition (SPOD) \citep{towne2018,schmidt2020} was employed to visualize the modal structures. The SPOD modes denoted by $\Psi_k$, $k=1, 2, ..., N$, where $N$ is the number of SPOD modes, have a corresponding eigenvalue $\lambda_1\ge\lambda_2, ..., \ge \lambda_N$, which is a measure of the total perturbation kinetic energy contained in each mode. These modes are orthogonal for given $f$ and $\beta$ and sorted in decreasing energy. Therefore, the first modes describe the leading structures contributing to the flow dynamics. Moreover, these modes evolve coherently in space and time \citep{towne2018}. In the current analysis, 2,000 snapshots spaced with a time step $\Delta t=5\times10^{-3}$ are divided into ten blocks with $80\%$ overlap. The convergence of the method was evaluated by comparing the leading SPOD modes obtained with 75\% and the full-time series, indicating minimal deviations.

\begin{figure}
    \subfigure[$AoA=12.8^\circ$, $f=16$, $Ro_r=3.1$.
    \label{fig:SPOD_L1_o045t_f16_b0}]
    {\includegraphics[width=0.495\linewidth]
        {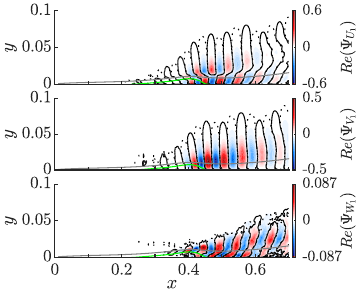}}%
    \subfigure[$AoA=12.8^\circ$, $f=16$, $Ro_r=0$.
    \label{fig:SPOD_L1_o045n_f16_b0}]
    {\includegraphics[width=0.495\linewidth]
        {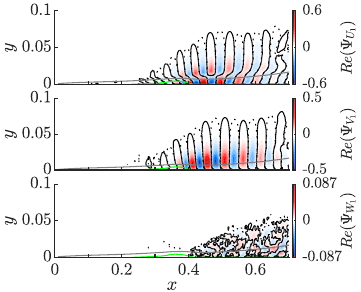}}

    \subfigure[$AoA=4.2^\circ$, $f=10$, $Ro_r=6.3$.
    \label{fig:SPOD_L1_o09t_f10_b0}]
    {\includegraphics[width=0.495\linewidth]
        {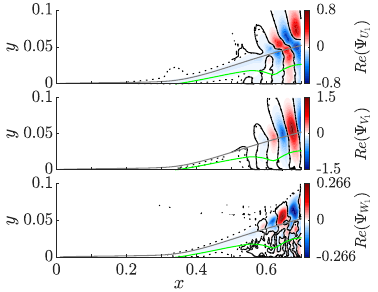}}%
    \subfigure[$AoA=4.2^\circ$, $f=10$, $Ro_r=0$.
    \label{fig:SPOD_L1_o09n2_f10_b0}]
    {\includegraphics[width=0.495\linewidth]
        {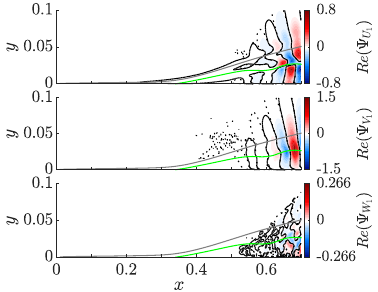}}%

   \subfigure[$AoA=1.2^\circ$, $f=10$, $Ro_r=9.4$.
    \label{fig:SPOD_L1_o135t_f10_b0}]
    {\includegraphics[width=0.495\linewidth]
        {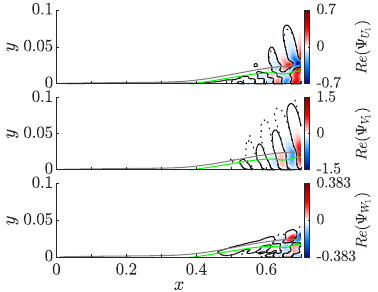}}%
    \subfigure[$AoA=1.2^\circ$, $f=10$, $Ro_r=0$.
    \label{fig:SPOD_L1_o135n_f10_b0}]
    {\includegraphics[width=0.495\linewidth]
        {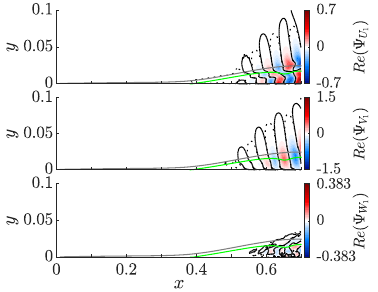}}%        
    \caption{Real part of the first SPOD mode of streamwise, normal, and spanwise velocity perturbations for $\beta=0$. \protect\blackline and \protect\dottedblackline indicate contour levels 0.001 and -0.001. \protect\grayline is the displacement thickness ($\delta^*$). \protect\greenline is the edge of the LSB.}
    \label{fig:SPOD_b0}    
\end{figure}

The first SPOD mode for frequencies close to the most amplified TS/KH instability is presented in Fig. \ref{fig:SPOD_b0}. For $AoA=12.8^\circ$, the wavepacket initiates just downstream of the separation location. A phase shift in the mode of streamwise velocity perturbation occurs at $y\approx\delta^*$, which separates the near-wall region with viscous instability from the inviscid instability area further into the free-stream. The normal velocity perturbation mode presents its peak amplitude at $y=\delta^*$. Overall the modes for streamwise and normal velocity perturbations are very similar in the rotating and non-rotating cases. However, the spanwise velocity perturbation mode presents a coherent wavepacket in the rotating case, starting at the front part of the LSB. The peak amplitude of this mode occurs at the location of the inflection point of the mean spanwise velocity profiles, situated above $y=\delta^*$. The instability engendered by this inflection point contributes little to the total perturbation kinetic energy, which can also be seen by the low relative energy of $Re(\Psi_{W_1})$ compared to the other components. In the non-rotating case, this coherent structure is not present. Regarding $AoA=4.2^\circ$, the TS/KH mode wavepacket is also formed. Nonetheless, the mode lies farther away from the wall due to the thick shear layer and tends to a pure KH instability from a free shear layer. While the peak amplitudes occur at $y=\delta^*$ in the rotating case, this occurs close to the LSB edge in the non-rotating one. Notice that this wavepacket is only formed downstream of where $y_{in}/y_{max_{\mathcal{P}}} \rightarrow 1$ for the mean streamwise velocity profile, clearly indicating its inviscid character. The $Re(\Psi_{W_1})$ mode is also coherent in the rotating case with peak amplitude at the location of the inflection point in the mean spanwise velocity profiles. Moreover, the amplitude of this mode is higher than in the $AoA=12.8^\circ$ case. The non-rotating case does not display coherent $Re(\Psi_{W_1})$. Finally, the $AoA=1.2^\circ$ case presents a mode that is qualitatively similar between the rotating and non-rotating cases, considering $Re(\Psi_{U_1})$ and $Re(\Psi_{V_1})$, also corresponding to a TS/KH instability. Regarding $Re(\Psi_{W_1})$, even though apparently less coherent, $Re(\Psi_{W_1})$ presents higher amplitude than for higher $AoA$. The coherence of this mode increases with frequency.

%-------------------------------------------------------------------------------
%------------------------------------------------------------------------------

\begin{figure}
    \subfigure[Traveling crossflow mode for $f=2$.
    \label{fig:SPOD_L1_o09t_f2_b188.5}]
    {\includegraphics[width=1\linewidth]
        {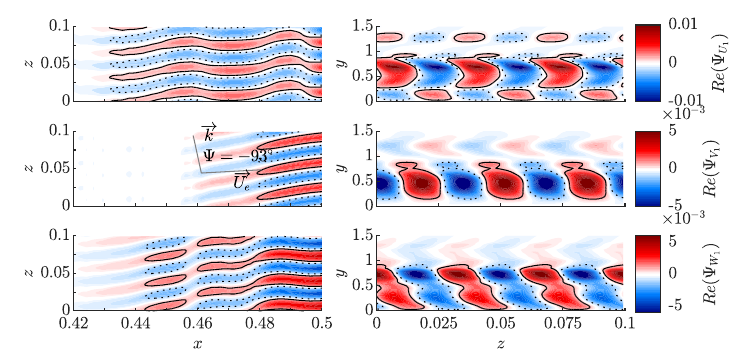}}

   \subfigure[Stationary crossflow mode.
    \label{fig:SPOD_L1_o09t_f0_b188.5}]
    {\includegraphics[width=1\linewidth]
        {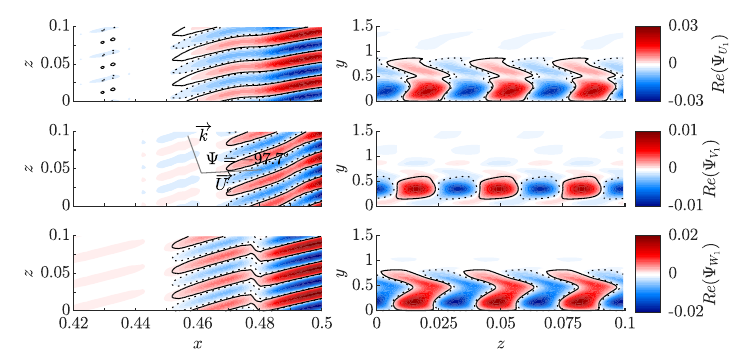}}%
    \caption{Real part of the first SPOD mode of streamwise, normal, and spanwise velocity perturbations for $\beta=188.5$, $AoA=4.2^\circ$, $Ro_r=6.3$ at $x=0.49$ and $y=0.0045$ ($0.2\delta^*$). The isocontours of amplitude 0.001 and -0.001 are marked with \protect\blackline and \protect\dottedblackline, respectively. $\overrightarrow {k}$ is the wavevector and $\overrightarrow {U_e}$ is the velocity vector at the inviscid streamline.}
    \label{fig:SPOD_b188.5}    
\end{figure}

Figure \ref{fig:SPOD_b188.5} shows the first SPOD modes for the $AoA=4.2^\circ$ case in rotating condition for $\beta=188.5$ ($n_z=3$). Figure \ref{fig:SPOD_L1_o09t_f2_b188.5} shows the results for $f=2$, which presents the characteristics of a traveling crossflow mode with vortices nearly aligned with the streamwise direction on a plane parallel to the airfoil (left panels). $\Psi=-93^\circ$, as shown in the panel for $Re(\Psi_{V_1})$, in the range of crossflow modes. The contours on the spanwise plane (right panels) indicate that the SPOD mode consists of counter-rotating vortices with peak amplitude at $y/\delta^*\approx0.7$ for $Re(\Psi_{U_1})$ and $Re(\Psi_{W_1})$ and $y/\delta^*\approx0.5$ for $Re(\Psi_{V_1})$), the same locations predicted by linear stability analysis. This further indicates that these are crossflow vortices. Notice that their energy is much lower than that of the maximum amplified TS/KH waves, suggesting that they do not act as the primary instability driver but rather influence the secondary instability process. The SPOD modes for $f=0$, portrayed in Fig. \ref{fig:SPOD_L1_o09t_f0_b188.5}, display an angle of $\Psi=-97.7^\circ$ relative to the inviscid streamline, which suggests these are stationary crossflow modes. The amplitude of these modes is higher than that of stationary crossflow vortices. It increases with $x$ but is probably insufficient for causing transition alone since the transition location occurs at $x=0.5$. The results on the spanwise plane do not indicate so clearly a crossflow mode such as in the $f=2$ case since the maxima of $Re(\Psi_{U_1})$ and $Re(\Psi_{W_1})$ occur closer to the wall than expected. This can be due to the non-linear forcing of these modes by the TS/KH instability. Nevertheless, there are still energetic structures for these modes at $y/\delta^*\approx0.7$, where stability analysis indicates the maximum amplitude takes place. The modes found here were only present in the rotating case, proving they are engendered by rotation. Traveling crossflow vortices were also observed for $AoA=12.8^\circ$ and $AoA=1.2^\circ$ in the rotating condition.

%-------------------------------------------------------------------------------
%------------------------------------------------------------------------------

\section{Conclusions}\label{sec:conclusion}

Direct numerical simulations (DNS) of the flow over a section of a rotating wind turbine blade are performed to study the effects of rotation on the flow stability and laminar-turbulent transition on the suction side. The computations were carried out in the rotating frame of reference, including centrifugal and Coriolis forces as source terms in the momentum equations. The angles of attack ($AoA$) of $12.8^\circ$, $4.2^\circ$, and $1.2^\circ$ were selected to assess the role of different flow separation conditions, which attempts to overcome the shortcomings pointed out by \citet{guntur2012} in the study by \citet{gross2012}. Three accompanying simulations with the same $AoA$ were also conducted without incorporating centrifugal and Coriolis force and with zero spanwise velocity to emulate corresponding non-rotating cases. All simulations considered a chord Reynolds number of $1 \times 10^5$. The results were analyzed using Fourier analysis and spectral proper orthogonal decomposition (SPOD). Moreover, primary stability analyses were performed to assess the roles of convective and absolute instabilities.

The effects of rotation depend on the difference between the local and free-stream velocities. Rotation accelerates the attached flow downstream in regions under a strong adverse pressure gradient (APG), counteracting the latter. On the other hand, in the case of a streamwise accelerating flow, i.e., under a favorable pressure gradient (FPG), rotation decelerates the flow, acting as an APG. Nevertheless, the effects are the opposite when reversal flow occurs, such as in laminar separation bubbles (LSBs). Upon sufficiently high reverse flow, the Coriolis force further decelerates the stream, increasing the LSB compared to the non-rotating case. These results are valid when the spanwise velocity is positive towards the root. One may expect the reverse of them under the occurrence of tip flow.

The $AoA=12.8^\circ$ case presents a strong (APG) since the leading edge and the rotation effects tend to counteract that by accelerating the flow downstream prior to separation. Inside the LSB, rotation strongly decelerates the flow increasing the reverse flow from -7\% in the non-rotating case to -16\% in the rotating one. Separation occurs at the same location in the rotating and non-rotating cases, but reattachment occurs 4\% more downstream in the rotating condition, and the LSB height is 74.7\% higher. These rotation-induced flow modifications stabilize the boundary layer upstream of separation concerning Tollmien-Schlichting (TS) waves. Nevertheless, inside the LSB, particularly near the maximum LSB height, rotation increases the growth rates of the mixed Tollmien-Schlichting/Kelvin-Helmholtz (TS/KH) instability that develops in the separated shear layer. The secondary instability of these rolls occurs via Marxen's oblique mechanism \citep{marxen2003} involving the triad $(1,0)$, $(0,\pm 2)$, and $(1,\pm 2)$, followed by a sub-harmonic growth of $(0.5,\pm 1)$ modes. Both mechanisms are enhanced in the rotating case, which leads to a rapid breakdown of the rolls to small-scale turbulence. This is characteristic of an oblique mechanism \citep{rist_vki}. On the other hand, the non-rotating case displays a predominantly sub-harmonic secondary instability, with the dilation of the streamwise wavelength, which leads to a slower breakdown of the two-dimensional structures. Another effect of rotation is rendering the above modes with $\beta<0$ dominant in the secondary instability process. Due to the stabilization of the attached boundary layer upstream and the increase in the reverse flow, transition is 3\% more downstream in the rotating case. The spanwise flow in the rotating case does not enhance primary instabilities, albeit coherent structures in the spanwise velocity appear in the TS/KH wavepacket. Reattachment occurs 4\% more downstream in the rotating case.

Considering the $AoA=4.2^\circ$ and $AoA=1.2^\circ$ cases, due to the prolonged region with FPG upstream of separation, the flow is decelerated there, rendering TS waves more unstable. However, this trend is reversed after separation, stabilizing the mixed TS/KH mode in the separated shear layer. Due to the higher rotation rate, a flow directed towards the blade tip appears in the fore part of the LSB. These crossflow velocity profiles are inflectional, and stationary crossflow modes emerge in the $AoA=4.2^\circ$ case and traveling ones in both $AoA=4.2^\circ$ and $AoA=1.2^\circ$ cases, rendering the flow more unstable. Nonetheless, the occurrence of radial flow may reduce the growth rates of these modes, similar to the observed for a rotating disk with axial inflow \citep{hussain2011,deschamps2017}. Therefore, the low growth rates of these modes do not compensate for the stabilization of the TS/KH mode by rotation inside the LSB. Despite a higher initial growth in the rotating case, the amplitude and growth rates of the primary instability are higher in the non-rotating one close to transition. In both cases, the streamwise vortices $(0,\pm 1)$ and $(0,\pm 2)$ present a high amplitude. This is particularly true in the non-rotating case for $AoA=4.2^\circ$, where these modes modulate the spanwise rolls and lead to their breakdown. The rotating case presents lower amplitudes of these modes, and the spanwise modulation of the two-dimensional rolls is less visible. The sub-harmonic modes $(0.5,-1)$ is more dominant in the rotating case, representing a change in the trend noticed for $AoA=12.8^\circ$, where the sub-harmonic mechanism was more relevant in the non-rotating case and the oblique one in the rotating condition. The secondary instability in the $AoA=1.2^\circ$ case is predominantly sub-harmonic in both conditions but with the predominance of $\beta<0$ modes in the rotating case. The spanwise flow sets a preferential direction for disturbance growth and leads to a lack of symmetry around the $\beta=0$ axis. The coherent structure in the spanwise velocity formed together with the TS/KH mode is stronger for $AoA=1.2^\circ$, probably due to the higher rotation rate. However, no change in the transition location is noticed between rotating and non-rotating conditions for $AoA=4.2^\circ$ and $AoA=1.2^\circ$, when measured as the location of the maximum boundary-layer shape factor. A finite region of absolute instability exists between the first two recirculation cells in these cases. This may indicate the global instability of the flow \citep{huerre1990}. Rotation rendered the absolute instabilities less unstable in the $AoA=4.2^\circ$ case, possibly due to the streamwise acceleration of the flow inside the LSB. For the $AoA=1.2^\circ$ case, this effect also occurred in the first recirculation cell, but rotation enhanced the absolute instability in the second one due to a stronger reverse flow.

% \appendix

%%%%%%%%%%%%%%%%%%%%%%%%%%%%%%%%%%%%%%%%%%%%%%%%%%%%%%%%%%%%%%%%%%%%%%%%%%%%%%%%%
%%%%%%%%%%%%%%%%%%%%      ACKNOWLEDGEMENTS  %%%%%%%%%%%%%%%%%%%%%%%%%%%%%%%%%%%%%
%%%%%%%%%%%%%%%%%%%%%%%%%%%%%%%%%%%%%%%%%%%%%%%%%%%%%%%%%%%%%%%%%%%%%%%%%%%%%%%%%

%%\section*{Acknowledgments}

\begin{acknowledgements}
This research was supported by StandUp for Energy and StandUp for Wind. The computations were performed on resources from the Swedish National Infrastructure for Computing (SNIC) at the PDC Center for High-Performance Computing at the Royal Institute of Technology (KTH), the National Supercomputer Centre at Link{\"o}ping University, and the High-Performance Computing Center North (HPC2N) at Umeå University.
\end{acknowledgements}

\section*{Compliance with Ethical Standards}

The authors declares that they have no conflict of interest.

%--------------------------------------------------------------------------------
%\clearpage

%\section*{References}

\markboth{Bibliografia}{REFERENCES}
\bibliographystyle{elsarticle-num-names}

%\bibliography{references2}

%%%%%%%%%%%%%%%%%%%%%%%%%%%%%%%%%%%%%%%%%%%%%%%%%%%%%%%%%%%%%%%%%%%%%%%%%%%%%%%%%
%%%%%%%%%%%%%%%%%%%%%%%%%%%%%%%%%%%%%%%%%%%%%%%%%%%%%%%%%%%%%%%%%%%%%%%%%%%%%%%%%
%%%%%%%%%%%%%%%%%%%%%%%%%%%%%%%%%%%%%%%%%%%%%%%%%%%%%%%%%%%%%%%%%%%%%%%%%%%%%%%%%
\end{document}